\documentclass{emulateapj}

\newcommand{\msun}{M$_{\odot}$}

\slugcomment{}
\usepackage{graphicx}
\usepackage{longtable}
\usepackage{subfigure}
\usepackage{amssymb, amsfonts}
\usepackage{amsmath}
\usepackage{booktabs}
\usepackage{tablefootnote}

\shorttitle{How Do Galaxies Trace a Large Scale Structure?}
\shortauthors{Shi et al.}

\begin{document}

\title{How Do Galaxies trace a large scale structure?: \\A case study around a massive protocluster at $z=3.13$}

\author{Ke Shi\altaffilmark{1}, Yun Huang\altaffilmark{1}, Kyoung-Soo Lee\altaffilmark{1,2}, Jun Toshikawa\altaffilmark{3}, Kathryn N. Bowen\altaffilmark{1}, Nicola Malavasi\altaffilmark{4},\\ B. C. Lemaux\altaffilmark{5,8}, Olga Cucciati\altaffilmark{6,7}, Olivier Le Fevre\altaffilmark{8}, Arjun Dey\altaffilmark{9}}
\altaffiltext{1}{Department of Physics and Astronomy, Purdue University, 525 Northwestern Avenue, West Lafayette, IN 47907}
\altaffiltext{2}{Visiting astronomer, Kitt Peak National Observatory (KPNO), National Optical Astronomy Observatory, which is operated by the Association of Universities for Research in Astronomy (AURA) under a cooperative agreement with the National Science Foundation.}
\altaffiltext{3}{Institute for Cosmic Ray Research,
the University of Tokyo, ICRR, 5-1-5 Kashiwanoha, Kashiwa, 277-0882 Japan.}
\altaffiltext{4}{Institut d'Astrophysique Spatiale, CNRS (UMR 8617), Universit$\acute{\rm e}$ Paris-Sud, B$\hat{\rm a}$timent 121, Orsay, France.}
\altaffiltext{5}{Department of Physics, University of California, Davis, One Shields Ave., Davis, CA 95616.}
\altaffiltext{6}{INAF--Osservatorio Astronomico di Bologna, via Gobetti, 93/3, I-40129, Bologna, Italy.}
\altaffiltext{7}{University of Bologna, Department of Physics and Astronomy (DIFA), V.le Berti Pichat, 6/2 - 40127, Bologna, Italy.}
\altaffiltext{8}{Aix Marseille Universit\'e, CNRS, LAM (Laboratoire d'Astrophysique de Marseille) UMR 7326, 13388, Marseille, France. }
\altaffiltext{9}{National Optical Astronomy Observatory, Tucson, AZ
85726.}



\begin{abstract}
In the hierarchical theory of galaxy formation, a galaxy overdensity is a hallmark of a massive cosmic structure. However, it is less well understood how different types of galaxies trace the underlying large-scale structure. Motivated by the discovery of a  $z=3.13$ protocluster, we examine how the same structure is populated by  Ly$\alpha$-emitting galaxies (LAEs). To this end, we have undertaken a deep narrow-band imaging survey sampling Ly$\alpha$ emission at this redshift. Of the 93 LAE candidates within a 36\arcmin$\times$36\arcmin~(70$\times$70~Mpc$^2$)  field, 21 galaxies form a significant surface overdensity ($\delta_{\Sigma,{\rm LAE}}=3.3\pm0.9$), which is spatially segregated from the Lyman break galaxy (LBG) overdensity.  One possible interpretation is that they  trace two separate structures of comparable masses ($\approx 10^{15}M_\odot$) where the latter is hosted by a halo assembled at an earlier time. We speculate that the dearth of LAEs in the LBG overdensity region may signal the role of halo assembly bias in galaxy formation, which would suggest that different search techniques may be biased accordingly to the formation age or dynamical state of the host halo. The median Ly$\alpha$- and UV luminosity is 30--70\% higher for the protocluster LAEs relative to the field. This difference cannot be  explained by the galaxy overdensity alone, and may require a top-heavy mass function, higher star formation efficiency for protocluster halos, or  suppression of galaxy formation in low-mass halos. A luminous Ly$\alpha$ blob and an ultramassive galaxy found in this region paint a picture consistent with the expected early growth of galaxies in clusters. 
\end{abstract}




\section{Introduction} \label{sec:intro}

In the hierarchical theory of structure formation, initial small density fluctuations give rise to the formation of first stars and galaxies. These structures subsequently grow larger and more massive via mergers and accretion \citep{White78}. In this context, galaxy clusters provide unique laboratories to study how galaxy formation proceeded in the densest cosmic structures. In the local universe, cluster galaxies form a tight `red sequence' \citep{Visvanathan77,Bower92} and obey the `morphology-density' relation \citep{Dressler80,Goto03}, showcasing the impact of dense environments on the star formation activities of the inhabitants. In addition, existing studies strongly suggest that cluster galaxies experienced early growth at an accelerated pace followed by swift shutdown of their star formation, and have been evolving passively in the last $\approx$~10~Gyr \citep[e.g.,][]{steidel05, Eisenhardt08, Hatch11, Koyama13, Cooke14, Husband16, Shimakawa17}.

The presence of massive quiescent galaxies in clusters out to $z\sim 1$ argues that the negative impact of dense environments must become less pervasive at earlier times, and that  the star formation-density relation may even reverse \citep[e.g.,][]{Elbaz07, Cooper08, Tran10, Koyama13, Brodwin13, Alberts14, Santos14, Welikala16} although it is still a matter of debate when this reversal occurs \citep[e.g., see][]{lemaux18b}. While the enhanced level of star formation activity in young forming clusters would certainly be consistent with the general expectations of cluster formation, direct evidence of this observational picture needs to come from distant galaxies residing in `protoclusters' at $z>2$, the epoch in which much of star formation activity and subsequent quenching are expected to have occurred.

Young protoclusters are far from virialized, and are distributed over large cosmic volumes with their angular sizes expected to span 10$\arcmin$--30$\arcmin$ in the sky \citep[e.g.,][]{Chiang13, Muldrew15}.  Moreover, the largest structures (those which will evolve into systems similar to Coma with their final masses exceeding $\gtrsim 10^{15}M_\odot$) are extremely rare with a comoving space density of $\approx$2$\times10^{-7}$~Mpc$^{-3}$ \citep{Chiang13}. Combined with their optical faintness, these characteristics make it observationally challenging to robustly identify protoclusters, and to conduct a complete census of their constituents  for those confirmed. 

Nevertheless, some protoclusters have been confirmed thanks to deep extensive spectroscopy of `blank fields' \citep[e.g.,][]{Steidel98, Steidel00,steidel05, Cucciati14, Lee14, Dey16, Lemaux14, wang16, Cucciati18, Jiang18}, which give us a glimpse of diverse galaxy types residing in protoclusters, such as luminous Ly$\alpha$ nebulae, dusty star-forming galaxies, and massive and quiescent galaxies. Studying these galaxies in details  will ultimately lead us to a deeper understanding of how cluster elliptical galaxies and the brightest cluster galaxy (BCG) are assembled. 

Another critical avenue in understanding cluster formation  is a detailed characterization of their large-scale environments. Such information will pave the way to understand how galaxies' star formation activity is linked to their immediate local density. One efficient way to do so is to pre-select candidate galaxies in overdense regions photometrically and follow them up with spectroscopy. Given the expected high star formation activity, a  selection of star-forming galaxies (such as Lyman break galaxies; LBGs hereafter) can provide a reasonable candidate pool, albeit not a complete one \citep[e.g.,][]{Toshikawa12, Lee14, Dey16, Toshikawa16,Toshikawa17}, from which possible overdense structures may reveal themselves as higher surface density regions \citep{Chiang13}. Alternatively, a narrow-band imaging selection sampling strong emission lines such as Ly$\alpha$ or H$\alpha$ has emerged as a popular choice as it allows sampling of a small slice of cosmic volume. Such emission-line based selection methods are  advantageous in defining environments with minimal contamination from fore- and background interlopers \citep[e.g.,][]{Pentericci00, Venemans07, Overzier08, Kuiper11, Hatch11, Mawatari12, Cooke14,Yang10,Badescu17,Higuchi18}. 

Given that a galaxy overdensity is a hallmark of massive cosmic structures, any method that is able to detect them should, in principle, serve us equally well in identifying progenitors of massive clusters provided that their galaxy biases are well understood. Understanding how different galaxy populations trace the underlying large-scale structure -- not only LBGs and Ly$\alpha$ emitters (LAEs) but also other types such as AGN and dusty star-forming galaxies  that have been reported to reside in abundance in dense protocluster environments -- can illuminate the early stages of cluster elliptical formation,  and also help us fine-tune the search techniques in the future in the era of wide-area surveys such as Large Synoptic Survey Telescope and Hobby-Eberly Dark Energy Experiment. 

In this paper, we present a follow-up study of a galaxy overdensity in the D1 field of the Canada-France-Hawaii-Telescope Legacy Survey (CFHTLS). The structure  `D1UD01' was discovered as a result of a systematic search of protoclusters conducted by  \cite{Toshikawa16} where candidate structures were identified  based on their prominent surface densities of LBGs at $z\sim 3-5$. Follow-up spectroscopy confirmed five galaxies at $z=3.13$ located within 1~Mpc of one another, suggesting the possible existence of a highly overdense structure.  At this redshift, Ly$\alpha$ emission is conveniently redshifted into a zero-redshift [O~{\sc iii}] filter, providing us the unique opportunity to explore how line-emitting galaxies are populated in a massive structure identified and characterized by an independent method.

This paper is organized as follows.  In \S~2, we present the new narrow-band imaging of a subsection of the CFHTLS D1 field containing a confirmed protocluster at $z=3.13$. Combining the new observations with existing broad-band data, we identify a sample of LBGs and LAEs, and  conduct a search for Ly$\alpha$ nebulae in the field (\S~3). In \S~4, we measure their  angular distributions and identify possible overdensity regions.  In \S~5, we discuss the masses of their descendants,  examine a possible trend of star formation activity with local environment, and speculate the implications based on these results. A search for a proto-BCG is also presented. Finally, a summary of our results is given in \S~6. 

We use the WMAP7 cosmology $(\Omega, \Omega_\Lambda, \sigma_8, h) = (0.27, 0.73, 0.8, 0.7)$ from \citet{wmap7}. Distance scales are given in comoving units unless noted otherwise. All magnitudes are given in the AB system \citep{oke83}. In the adopted cosmology, 1\arcsec\ corresponds to the angular scale of 7.84 kpc at $z=3.13$.

\section{Data and photometry} \label{sec:data}

\subsection{New observations}

In September 2017, we obtain narrowband imaging of the protocluster candidate `D1UD01' and the surrounding region in the D1 field, one of the four Canada-France-Hawaii-Telescope Legacy Survey deep fields. The pointing center is [$\alpha$, $\delta$]=[36.316$^\circ$, $-4.493^\circ$]. The data are taken with the Mosaic 3 Camera \citep{mosaic3} on the Mayall 4m telescope of the Kitt Peak National Observatory (NOAO Program ID: 2017B-0087). The KPNO Mosaic [O~{\sc iii}] filter no.~k1014 ($o3$ filter, hereafter) is used, with a central wavelength of 5024.9\AA\ and a full-width-at-half-maximum (FWHM) of 55.6\AA. The $o3$ filter samples redshifted Ly$\alpha$ line in the range $z=3.132\pm0.023$, spanning a line-of-sight distance of 44~Mpc. 

The individual exposure time of 1200~sec is used with small-offset dithers ({\tt FILLGAP}) optimized to fill in CCD chip gaps. We discard the frames taken with seeing $>1\farcs3$. We identify and remove a handful of frames which appear to have been taken when the guide star was temporarily lost, resulting in the sources to leave visible trails in the image. The total exposure time of the new imaging is 14.0 hr. The mosaic image has a native pixel scale of 0.25$\arcsec$.

We calibrate the astrometry with the IRAF task {\tt msccmatch} using the stars identified in the CFHTLS deep survey catalog  \citep[][]{Gwyn12}, and re-project each image with a pixel scale of $0\farcs186$ using the tangent point of the CFHTLS images. The relative intensity scale is determined using the IRAF task {\tt mscimatch}. The reprojected frames are then combined into a final image stack using a weighted average, with the average weight inversely proportional to the variance of the sky noise measured in the reprojected frames. We trim the images removing the area near the edges with less than 20\% of the maximum exposure time, and mask areas near bright saturated stars. The final mosaic has an effective area of 0.32 deg$^2$ with a measured seeing of 1$\arcsec$.2.

As most of our observations were taken in non-photometric conditions, we calibrate the  photometric zeropoint using the CFHTLS broad-band catalogs. The central wavelength of the $g$ band is 4750\AA, reasonably close to that of the $o3$ filter at 5024.9\AA. We define a sample of galaxies that have the $g$-band magnitude  of 21--25~mag with the blue $g-r$ colors ($g-r\leq 0.2$), and determine the $o3$ band zeropoint such that the median $o3-g$ color is zero. We further check our result by plotting the $g-r$ colors vs $o3-g$ colors for all photometric sources. We confirm that the intercept in the $o3-g$ colors is zero. 

\begin{deluxetable}{cccc}[h]
\tablecaption{Data Set \label{table1}}
\tablehead{
\colhead{Band} & \colhead{Instrument} & \colhead{Limiting magnitude\tablenotemark{a}} & \colhead{FWHM} \\
\colhead{} & \colhead{} & \colhead{(5$\sigma$,AB)} & \colhead{($\arcsec$)}
}
\startdata
$u$ & MegaCam/CFHT & 27.50 & 0.8\\
$g$ & MegaCam/CFHT & 27.82 & 0.8\\
$o3$ & Mosaic-3/Mayall & 25.21 & 1.2\\
$r$ & MegaCam/CFHT & 27.61 & 0.8\\
$i$ & MegaCam/CFHT & 27.10 & 0.8\\
$z$ & MegaCam/CFHT & 26.30 & 0.8
\enddata
\tablenotetext{a}{ 5$\sigma$ limiting magnitude measured in a 2$\arcsec$ diameter aperture. 
}
\end{deluxetable}

\begin{figure}[h]
\epsscale{1.2}
\plotone{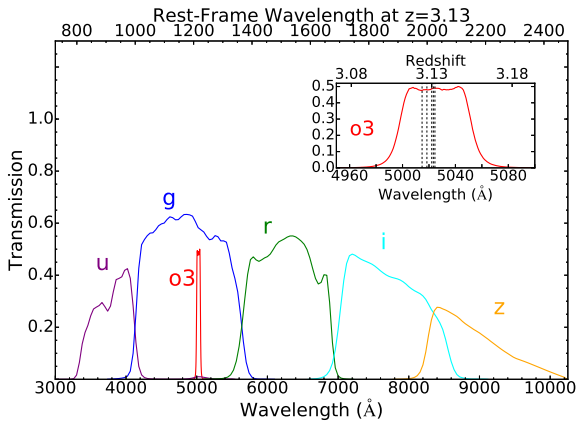}
\caption{
 Total throughput (filter+mirror+optics+CCD response) of the filters used to identify Ly$\alpha$ emitters in this work. The rest-frame wavelength range at $z=3.13$ is shown on the top axis. The inset zooms in on the $o3$ filter region. The corresponding redshift range of Ly$\alpha$ emission is indicated on top. The spectroscopic redshifts of the five confirmed members are indicated by dotted vertical lines.}  
\label{fig1}
\end{figure}

In conjunction with the new $o3$ data, we use the deep $ugriz$ images available from the CFHTLS Deep Survey  \citep{Gwyn12}. The broad band images are trimmed to have the identical dimension to the $o3$-band data. The photometric depth (measured from the sky fluctuations by placing 2\arcsec\ diameter apertures in random image positions) and native image quality of these bands are summarized in Table~\ref{table1}; their filter transmission curves are illustrated in Figure \ref{fig1}. 

\subsection{Photometry} \label{photometry}
We create a multiwavelength photometric catalog as follows.
First, we homogenize the PSFs of the broad-band data to match that of the worst-seeing data, i.e., the $o3$ image (FWHM=1.2\arcsec). The radial profile of the PSF in each image is approximated as a Moffat profile with the measured seeing FWHM, and a noiseless convolution kernel is derived using the IDL routine {\tt MAX$\_$ENTROPY}.  The broad band data is then convolved with their respective kernels to create a PSF-matched image. 

We create the narrow band catalog by running the SExtractor software \citep{Bertin96} in the dual image mode. The $o3$ band image is used for detection, while photometric measurements are performed in all the broad band images. The SExtractor parameter MAG\_AUTO is used to estimate the total magnitude, while colors are computed from the fluxes within a fixed isophotal area (i.e., FLUX\_ISO). As the images are PSF-matched, aperture correction in all bands is assumed to be given by the difference between MAG\_AUTO and MAG\_ISO estimated in the detection band.  A total of 43,940 sources are detected in the $o3$ image. We also use the broad-band-only  catalog released as part of the CFHTLS final data release\footnote{{\tt https://www.cfht.hawaii.edu/Science/CFHTLS/cfhtlsfinal\\releaseexecsummary.html}} (referred to as a `T0007' version, hereafter); the T0007 catalog contains 249,771 sources where   a $gri$ selected $\chi^2$ is used as a detection image.

\begin{figure*}
\epsscale{1.3}
\plotone{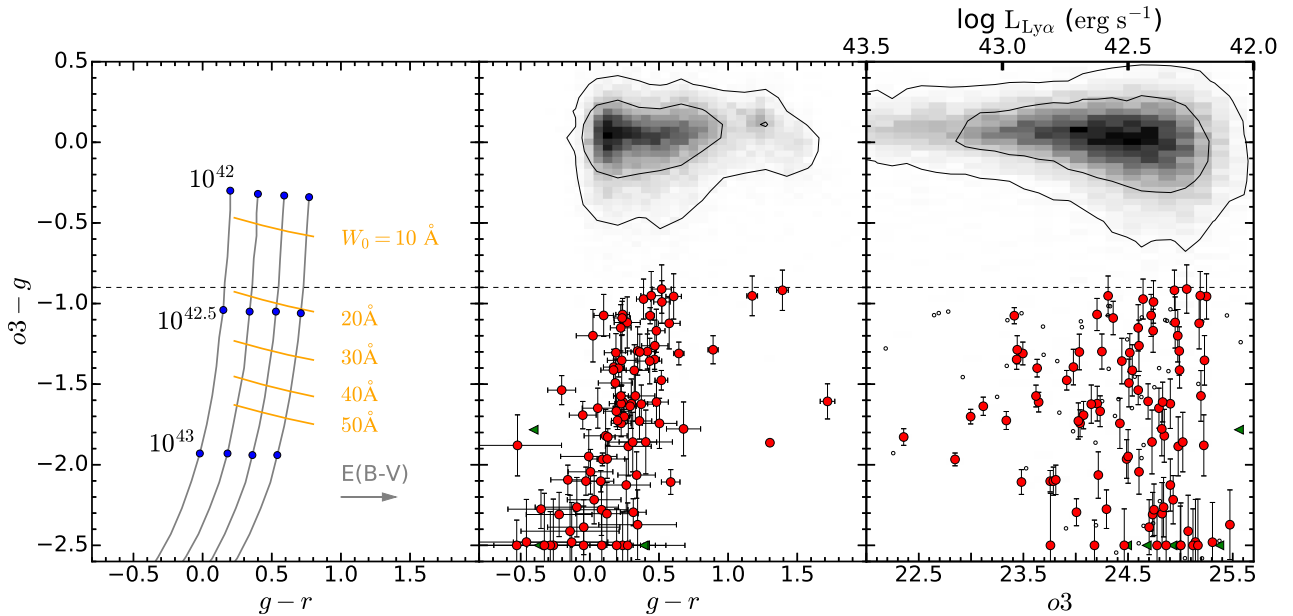}
\caption{
{\it Left:} Theoretical tracks for LAEs of different luminosities and dust reddening in the $o3-g$ vs $g-r$ color diagram. The grey lines show the color evolution with increasing Ly$\alpha$ luminosities (from top to bottom) at four  reddening values (${\rm E}(B-V) = 0-0.3$ in steps of 0.1 from left to right). Blue points show the Ly$\alpha$ luminosities, 10$^{42.0}$, 10$^{42.5}$ and 10$^{43.0}$ erg~s$^{-1}$ at the continuum $g$ band magnitude of 25.5. Orange lines represent the Ly$\alpha$ rest-frame equivalent widths ($W_0$) of 10, 20, 30, 40, and 50\AA. The $o3-g$ color cut (dashed horizontal line) approximately corresponds to $W_0 \gtrsim 20$\AA. {\it Middle:} the color-color diagram for all $o3$ detected sources. The two contour lines enclose 68\% and 95\% of the sources. Galaxies that satisfy the LAE criteria are indicated as red circles; those undetected in $g$ or $r$ band are shown as green triangles. Galaxies with $o3-g\leq -2.5$ are shown at the color position of $-2.5$. 
{\it Right:} $o3-g$ color as a function of $o3$-band magnitude. Sources that do not meet the $u-g$ color cut are shown in  open circles. The approximate Ly$\alpha$ luminosities corresponding to the $o3$ magnitude are indicated on the upper abscissa.}  
\label{fig:lae_selection}
\end{figure*}

\section{Analysis}\label{selection}

\subsection{Ly$\alpha$-emitting Galaxies at $z\sim3.13$}\label{subsec:lae}
The primary goal of this paper is to investigate the possible presence of a large scale structure in and around the five spectroscopic sources at $z=3.13$ discovered by \citet{Toshikawa16}. The $o3$ filter is ideally suited for this task as  redshifted Ly$\alpha$ emission  falls into it at $z=3.13\pm0.02$. The  redshift selection function, converted from the filter transmission, is illustrated in the inset of Figure~\ref{fig1}. The Ly$\alpha$-based spectroscopic redshifts of the five galaxies confirmed by \citet{Toshikawa16} are marked as vertical dashed lines. 

We adopt the following criteria to select LAE candidates at $z=3.13$: 
\begin{eqnarray}\label{lae_color}
o3-g<-0.9~~\wedge~~{\rm S/N}(o3)\geq 7 ~~\nonumber \\
\wedge~~[~u-g>1.2~~\vee~~{\rm S/N}(u)<2 ~]
\end{eqnarray}
where the symbols $\vee$ and $\wedge$ are the logical ``OR'' and ``AND'' operators, respectively, and S/N denotes the signal-to-noise ratio  within the isophotal area. The $u-g$ color criterion requires a strong continuum break falling between the two filters to ensure that the source lies at $z\gtrsim2.7$. 

To design the  selection criteria, we synthesize the colors by generating model galaxies spanning a  range of rest-frame UV continuum slope, Ly$\alpha$ emission line equivalent width (EW), and Ly$\alpha$ luminosity.  The galaxy's spectral energy distribution (SED) is constructed assuming a constant star formation history observed at the population age of 100~Myr, with a \citet{Salpeter55} initial mass function and solar metallicity.  We account for attenuation by intergalactic hydrogen using the H{\sc i} opacity given by \citet{madau95}, and assume that the interstellar extinction obeys the \citet{calzetti00} reddening law. 

To the reddened, redshifted galaxy SED,  we add a Ly$\alpha$ emission with a Gaussian line profile centered at $1215.67 (1+z)$\AA\ and an intrinsic line width of 3\AA. The redshift $z=3.13$ is assumed.  Given that the $o3$ filter is much wider than the line width, exact values assumed for the line width are not important as long as they reproduce the observed galaxy colors and line FWHM reasonably well. The Ly$\alpha$ limiting luminosity from the above criteria is  $\approx$ 10$^{42.3}$ erg~s$^{-1}$. No extinction is applied to the Ly$\alpha$ line as it represents the observed luminosity.

In the left panel of Figure~\ref{fig:lae_selection}, we show the expected $o3-g$ and $g-r$ colors for  different reddening values with different line luminosities. For the Ly$\alpha$ luminosities indicated in the same panel, we assume a continuum $g$ band magnitude of 25.5~mag, which is based on the median value of our LAE sample. Finally, we stress that our photometric criteria (Equation~\ref{lae_color}) are sensitive to the line equivalent width and redshifts of the source, but not to the choice of IMF and metallicity adopted to create the base galaxy SED. For example, if a sub-solar metallicity ($Z=0.008$) is assumed, the $g-r$ colors would be bluer by 0.04~mag while the $o3-g$ colors would remain unchanged. 

The middle and right panel of Figure~\ref{fig:lae_selection} show the color-color and color-magnitude distributions of the $o3$-detected sources.  The adopted selection criteria (Equation~\ref{lae_color})  correspond to the rest-frame equivalent widths $\gtrsim 20$\AA\ at the target redshift range, and result in 94 LAE candidates.  Their $g-r$ colors suggest that the majority are consistent with being relatively dust-free with  a few exceptions. 
The LAE candidates are distributed over a $\approx 4,365$ Mpc$^2$ (1156~arcmin$^2$) field. With the exception of six (green triangles in Fig~\ref{fig:lae_selection}), all have robust continuum detections in the $g$ or $r$ band. 

Based on the photometric data, we derive the physical properties of our LAE candidates including the rest-frame Ly$\alpha$ EW ($W_0$), Ly$\alpha$ luminosity ($L_{{\rm Ly}\alpha}$), UV continuum luminosity at the rest-frame 1700\AA\ ($L_{1700}$), and UV spectral slope ($\beta$: defined as $f_\lambda \propto \lambda^\beta$). The Ly$\alpha$ luminosity and EW are derived following the prescription given in \citet{xue17}, which fully takes into account the Ly$\alpha$ forest attenuation in the relevant filters. 
The UV slope is computed from a linear regression fitting of the $riz$ photometric data; the continuum luminosity $L_{1700}$ is then extrapolated from the $i$-band flux density assuming the slope $\beta$. 
These quantities are listed in Table~\ref{tbl2}.

Four galaxies in our LAE sample are significantly redder  ($g-r>1.0$) than the majority. We check them  in the image to verify these sources are real and robust detections. One is likely an AGN with an extremely high UV luminosity ($r=21.7$ mag) and a point-like morphology. The other three may be more dust reddened than the other 90 LAE candidates. Dusty LAEs are rare, but have been reported in the literature \citep{Oteo12,Bridge13}, some of which are IR-luminous galaxies detected in mid-infrared surveys.  Assuming the \citet{calzetti00} dust law, their UV slope $\beta$ values correspond to the color excess of the stellar continuum ${\rm E}(B-V)$ of 0.20, 0.16, and 0.23, respectively, compared to the median value of 0.10 for the full LAE sample. These values are comparable to those measured for dusty LAEs with {\it Herschel}/PACS detection studied by \citet{Oteo12}.  

Three of the five spectroscopic sources in the `D1UD01' structure  satisfy our LAE selection; their IDs in the \citet{Toshikawa16} study are D1UD01-8, -9 and -6. Their Ly$\alpha$ EWs estimated from spectroscopy are 7.8, 21.0, 81.5\AA, respectively.  The remaining two, Toshikawa source ID D1UD01-7 and  D1UD01-10, do not meet our LAE selection because they are too faint in the $o3$ band (S/N in the range of 4--5); however, their $o3-g$ colors, $-1.61\pm 0.10$ and $-1.33\pm 0.08$, are  consistent with the Ly$\alpha$ EWs, 36.2 and 34.3\AA, measured from spectroscopy. \\

\noindent{\it Sample Contamination ~~} 
At the central wavelength of the $o3$ filter (5024.9\AA), the only plausible contaminants of our photometric LAE sample are [O~{\sc ii}] emitters at $z\sim0.35$, since our survey samples an inadequately small volume for [O~{\sc iii}] emitters which would lie at $z\sim0.01$. The adopted $o3-g$ color cut corresponds to the observed line EW of 83\AA, much larger than the values measured for [O~{\sc ii}] emitters, which mostly range in $\lesssim 50$\AA\ \citep{Hogg98,Ciardullo13} at $z=0.35$. The requirement that the galaxies have red $u-g$ colors provides an additional assurance that the Lyman break falls in the $u$ band (i.e., the sources lie at $z>2.7$).

Low-luminosity AGN  with a broad Ly$\alpha$ emission line at $z>2.7$ can potentially contaminate our LAE sample although the contamination is expected to be generally low \citep[at $\sim$ 1\%:][]{Gawiser06,Ouchi08,Zheng10,Sobral18}.  We cross-correlate the source positions with the X-ray sources listed in the XMM survey in the field \citep{Chiappetti05}, and find no match. However, the brightest source in our sample (QSO30046, $o3$=21.06, $r$=21.69~mag) is detected in the {\it Spitzer} MIPS 24~$\mu$m data. QSO30046 is also observed by the VIMOS VLT Deep Survey \citep[VVDS:][]{Lefevre13} and classified as an AGN at $z_{\rm spec}=3.86$. Given its redshift, the blue $o3-g$ color is owing to broad emission from Ly$\beta$ and O~{\sc vi} \citep{vandenberk01}. While we list its properties in Table~\ref{tbl2}, we remove this source from our LAE catalog.

We also cross-match our LAEs with spectroscopic redshift sources published by \cite{Toshikawa16} and those in the VVDS and VIMOS Ultra Deep Survey \citep[VUDS:][]{Lefvre15}. Four matches are found; three are part of the LBG overdensity reported by \citet{Toshikawa16} and the fourth lies at $z=3.133$, but well outside it spatially.  The relatively low number of matches is not surprising given that all these spectroscopic surveys are limited to sample only relatively bright sources (i.e., $i<25$). In comparison, the mean $i$ band magnitude of our LAE sample is $\sim26$. Furthermore, the `D1UD01' region is excluded from the VUDS survey coverage.

\subsection{Selection of LBG candidates}\label{subsec:lbg}

\begin{figure*}[ht!]
\epsscale{1.1}
\plotone{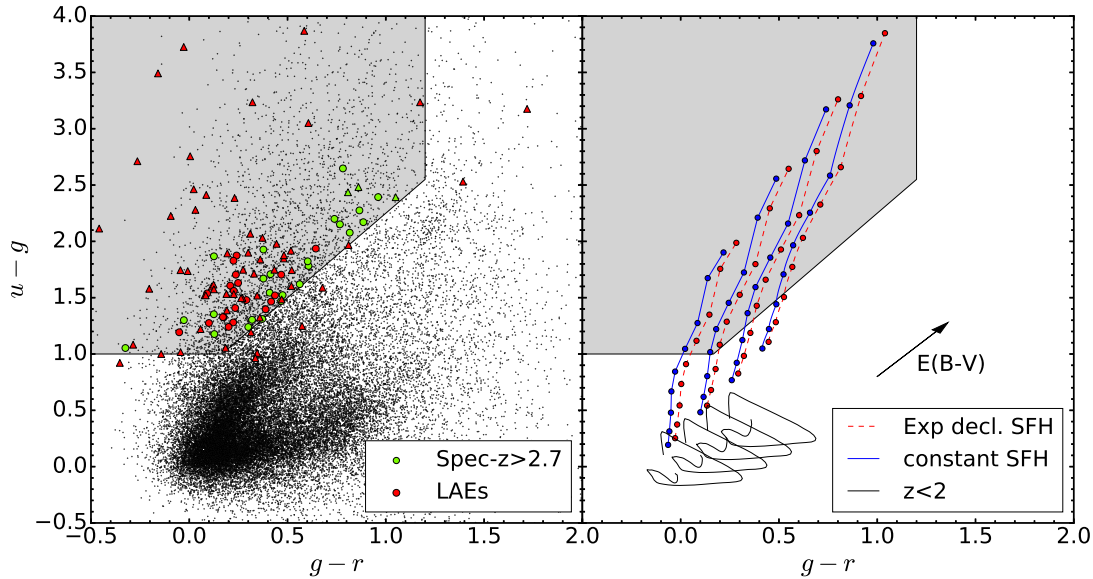}
\caption{
 {\it Left:} $u-g$ vs $g-r$ colors of all $r$-band detected sources are shown in dots together with the  LBG selection indicated by a grey shaded region. Red symbols are photometric LAEs while green symbols are known spectroscopic sources at $z_{\rm spec}\geq 2.7$. All sources that are not detected in the $u$-band are shown as triangles using the 2$\sigma$ limiting magnitude. {\it Right:} the redshift evolution of colors are illustrated for galaxies with dust reddening values. ${\rm E}(B-V)$=0, 0.1, 0.2, 0.3  (from left to right). Galaxy's star formation rate is modeled to be constant (blue) or declining exponentially with time (red) as $\psi\propto \exp{-(t/{\rm 100~Myr})}$. Along a given track, source redshift increases from $z=2.7$ upwards with the interval $\Delta z=0.1$. Black lines show expected colors of local spiral galaxies when redshifted out to $z=2$.
}
\label{fig:lbg_selection}
\end{figure*}

We also identify a sample of UV-luminous star-forming galaxies at $z\sim3$ by applying the Lyman  break color selection technique to the $ugr$ data from the CFHTLS T0007 catalog. The technique can identify star-forming galaxies with a modest amount of dust  by detecting spectral features produced by the Lyman limit at $\lambda_{\rm rest}=912$\AA\ and absorption by the intervening Ly$\alpha$ forest at $\lambda_{\rm rest}=912-1216$\AA. At $2.7<z<3.4$, both of these features  fall between the $u$ and $g$ bands. 

In the right panel of Figure~\ref{fig:lbg_selection}, we  show the expected redshift evolution of broad-band colors from $z=2.7$ in steps of $\Delta z=0.1$. Four reddening parameters are assumed, E($B-V$)=0.0, 0.1, 0.2, and 0.3 (from left to right). The synthetic colors of  lower-redshift galaxies are also computed using the \citet{Coleman80} template of S0 galaxies redshifted out to $z=2$ (black lines). As can be seen in the figure, most of the $z\sim3.1$ sources are located at $u-g>1.0$ while safely avoiding the locus of $z<2$ galaxies. Based on these considerations, we adopt the following criteria to select LBG candidates:
\begin{eqnarray}
u-g>1.0~~\wedge~-1.0<(g-r)<1.2 \nonumber~~ \\
\wedge~~(u-g) > 1.5 (g-r)+ 0.75~~~~~~~~~~
\end{eqnarray}
These are identical to those used by \citet{Toshikawa16}.

In the left panel of Figure~\ref{fig:lbg_selection}, we show the locations of all the sources in the two-color diagram.  For the sources undetected in the $u$ band, we show  lower limits by adopting the 2$\sigma$ limiting magnitude (28.5~mag).  We also require that the candidates   be detected with more than 3$\sigma$  ($7\sigma$) significance in the $g$ ($r$) bands to ensure that their detection and color measurements are robust.  A total of 6,913 galaxies are selected as our LBG candidates. 80 (86\%) of the LAEs  satisfy the adopted LBG criteria, with most of the remaining LAEs lying close to the selection criteria, confirming the similarity of the two populations. Our LBG catalog recovers  24  LBGs spectroscopically confirmed by \cite{Toshikawa16}  including all five `D1UD01' sources.
Of 6,913 galaxies, 210  have spectroscopic redshifts measured from the VVDS and VUDS surveys and by \citet{Toshikawa16}. Of those, 27 lie at $z<2.7$ yielding a contamination rate of 13\%.

The majority of these 27 galaxies have redshifts close to $z=2.7$, suggesting that they are simply scattered into the LBG window. To  quantify the role of photometric scatter, we carry out realistic galaxy simulations similar to those described in \citet{lee12a}. First, we create SEDs spanning a wide range of physical parameters (age, reddening, and redshift) and compute input photometry of these SEDs in the observed passbands. Mock galaxies are inserted into the images, and detection and photometric measurements are performed using the identical manner as the real data. The galaxies which satisfy our LBG criteria are collated into the master list. The redshift distribution of LBG-selected mock galaxies in the magnitude range $r=22-28$ (matching the optical brightness of our LBG sample) peaks at $z\sim3.1$ with a FWHM of $\sim 0.7$. Of those, 12\% lie at $z<2.7$, nearly identical to the contamination rate of 13\% estimated from spectroscopy. 

We make a qualitative comparison of the T0007 catalog with the \citet[][T16, hereafter]{Toshikawa16} catalog.
The major difference is a detection image which is a $gri$-based $\chi^2$ image for the T0007 catalog and the $i$-band for the T16 catalog. The detection setting (including the threshold) is also different. Overall, we find that the T16 catalog is more inclusive of fainter objects with the median $r$-band magnitude of 26.5~mag, compared to 26.0~mag for the T0007 catalog. The two catalogs have 4,219 sources in common, which accounts for 61\% and 54\% of the T0007 and T16 catalogs, respectively. 

\subsection{Search of Ly$\alpha$ Blobs}\label{sec:lab}

\begin{figure*}[ht!]
	\begin{subfigure}
		\centering
		\includegraphics[scale=0.7]{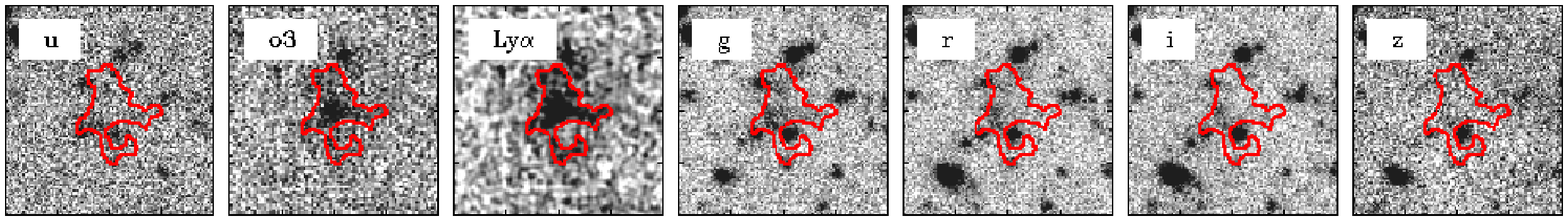}
		\end{subfigure}
			
		\begin{subfigure}
		\centering
		\includegraphics[scale=0.7]{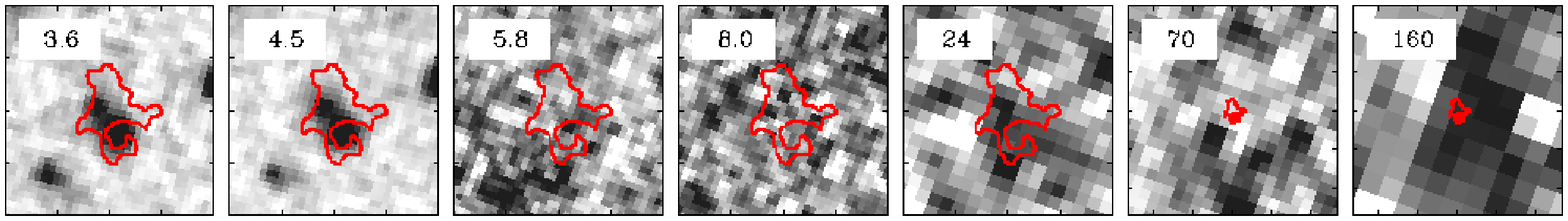}
		\end{subfigure}
\caption{Postage-stamp images of the LAB candidate LAB17139. In all panels, north is up and east is to the left. Each image is 20\arcsec\ on a side except for the 70~$\mu$m and 160~$\mu$m data which are 80\arcsec\ on a side. A red contour outlines the boundary of the Ly$\alpha$  isophote (see text).
}
\label{fig:lab_chart}
\end{figure*}

\begin{figure*}[ht]
\epsscale{1.1}
\plotone{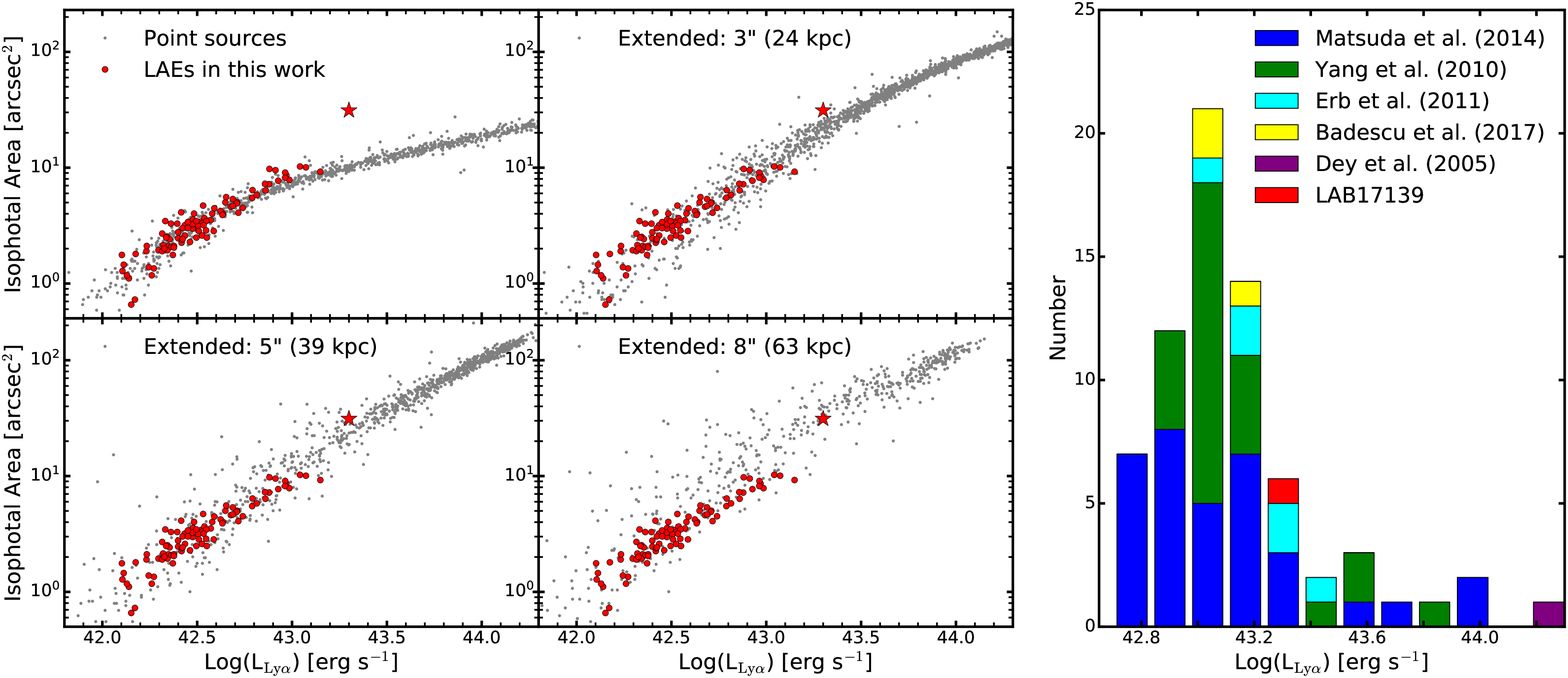}
\caption{
{\it Left:} the relationship between Isophotal size and intrinsic Ly$\alpha$ luminosity are determined through image simulations assuming that  the intrinsic light profile is point-like or falls off exponentially with a half-light radius of 3\arcsec, 5\arcsec, and 8\arcsec. Grey points show simulated galaxies while red circles indicate the measurements from the LAEs in this study. The Ly$\alpha$ nebula, LAB17139,  is marked as a red star in each panel. Given its observed luminosity and isophotal size, the half-light radius of LAB17139 is $\sim 40-55$~kpc (see text). 
{\it Right:} the distribution of Ly$\alpha$ luminosities of known giant Ly$\alpha$ nebulae at $z=2-4$ in the literature. LAB17139 is once again marked in red. 
}
\label{fig:lab_size}
\end{figure*}

We search for sources that are significantly extended in their Ly$\alpha$ emission; such sources are often referred to as a giant Ly$\alpha$ nebula or `Ly$\alpha$ blob' (LAB, hereafter). The largest LABs reported to date can be as large as $\gtrsim 100$~kpc across \citep[e.g.,][]{Dey05}. Multiple discoveries of luminous LABs  in and around galaxy overdensities \citep[e.g.,][]{Steidel00, matsuda04,Palunas04,Dey05,Prescott08,Yang10,Mawatari12,Badescu17} have led to a claim that they may be a signpost for massive large-scale structures. 

To enable a sensitive search, we first create a Ly$\alpha$ line image by estimating and subtracting out the continuum emission from the $o3$ image. Following the procedure described in \citet[][see their Eq.~11]{xue17}, the line flux is expressed as $F_{{\rm Ly}\alpha}=a f_{{\rm AB},o3} - b f_{{\rm AB},g}$, where $f$ is the monochromatic flux density in the respective bands, and $a$ and $b$ are coefficients that depend on the corresponding bandwidth and optical depth of the intergalactic medium as well as the UV continuum slope. For example, at $z=3.1$, with a UV slope $\beta$ of $-2.0$, $a\sim7.3\times10^{12}$ and $b\sim7.7\times10^{12}$.

We run the SExtractor software  \citep{Bertin96} on the Ly$\alpha$ image as a detection band and perform photometry on the $o3$ and $g$ band data. For detection, we require a minimum area of 16 pixels above the threshold $1.5\sigma$ which corresponds to 27.81 mag~arcsec$^{-2}$ or $1.80 \times 10^{-18}$~ergs~s$^{-1}$~cm$^{-2}$~arcsec$^{-2}$.  Our LAB search is slightly different from our LAE selection in that source detection is made on the Ly$\alpha$ image, and is tuned to be more sensitive to extended low surface-brightness sources. The same $o3-g$ color cut (Equation~\ref{lae_color}) as our LAE selection is applied. 
Our search yields a single  Ly$\alpha$ blob candidate in the entire field, which we name LAB17139. It is also identified as an LAE. At the Ly$\alpha$ luminosity of $\approx10^{43.3}$~erg~s$^{-1}$, it has the highest luminosity in our LAE sample. We estimate the isophotal area to be 31.2~arcsec$^2$ (1,920 kpc$^2$ assuming $z=3.13$). The postage-stamp images of LAB17139 are shown in Figure~\ref{fig:lab_chart}, and its properties are listed in Table \ref{tbl2}. 

At the centroid of its Ly$\alpha$ emission, no apparent counterpart exists in any of the broad band data ($gri$).  If its Ly$\alpha$ emission originates from a single galaxy, its continuum luminosity is fainter than $r$=28.6~mag ($2\sigma$). We do not find any plausible galaxy candidate in its vicinity that may lie at the same redshift. There are two UV bright sources just outside the isophote (one directly north and the other at the southwestern end); having the $u-g$ color of 0.57$\pm$0.19 and 0.90$\pm$0.12, neither of them satisfies our LBG selection. Therefore, it is unlikely they lie at the same redshift as LAB17139. 

We search for its possible infrared counterpart utilizing two publicly available {\it Spitzer} observations in the D1 field, namely the {\it Spitzer} Wide-area InfraRed Extragalactic survey \citep[SWIRE:][]{Lonsdale03} and the {\it Spitzer}  Extragalactic Representative Volume Survey \citep[SERVS:][]{Mauduit12}. The former includes all IRAC and MIPS bands while the latter was taken as part of post-cryogenic IRAC observations (3.6 and 4.5$\mu$m bands only, which are deeper than the SWIRE counterpart). In Figure~\ref{fig:lab_chart}, we show postage stamp images of these data centered on LAB17139. 

A single IR-bright source is identified within the LAB isophote which lies $\approx$1.2\arcsec\ away from the center of LAB17139; the source is securely detected in the 3.6 and 4.5$\mu$m bands and marginally detected in the MIPS 24$\mu$m, but not in the 70$\mu$m band. In the optical ($gri$) images, the source appears very diffuse and spans at least 2\arcsec. If it is a single source, it is likely  an interloper as it is too large to lie at $z\gtrsim3$. Given its clear positional offset from the centroid of the LAB, it is unlikely that the source is solely responsible for the Ly$\alpha$ emission. Thus, the physical association of this diffuse source and LAB17139 remains unclear.\\

\noindent {\it Intrinsic Size of LAB17139~~~~} We investigate the intrinsic size of LAB17139 by carrying out extensive image simulations. 
First, we insert artificial point sources with a range of luminosities into the Ly$\alpha$ image after convolving them with the image PSF, and recover them using the same detection setting as our LAB search.  On the top left panel of Figure~\ref{fig:lab_size}, we show how  measured isophotal size  correlates with luminosity for point sources (grey symbols). It is evident that the majority of our LAEs follow the same sequence except for a few highest luminosity LAEs. On the other hand, LAB17139 lies well above the point-source locus: i.e., its high luminosity is insufficient to explain its large size. 

Having established that the source is extended, we  repeat the simulation but  this time assuming that the radial profile of the source declines exponentially: $S(r)\propto \exp{[-1.6783 (r/r_s)]}$. In Figure~\ref{fig:lab_size}, we show the luminosity-isophotal area scaling relation for the sources with half-light radii of 3\arcsec\ ($r_s=1.8\arcsec$), 5\arcsec\ ($r_s=3.0\arcsec$), and 8\arcsec\ ($r_s=4.8\arcsec$); at $z=3.13$, these values correspond to 24, 39, and 63~kpc, respectively. At a fixed line luminosity, the scatter in the recovered isophotal area increases with sizes as expected due to lower surface brightness.  Nevertheless, Figure~\ref{fig:lab_size} shows that a unique scaling relation exists at a fixed intrinsic size. 

Utilizing this trend, with luminosities fixed in the simulation, we estimate that the half-light radius of LAB17139 must lie in the range of 39--55~kpc, provided that its surface brightness falls exponentially. Based on the average stack of 11 Ly$\alpha$ blobs at $z=2.65$, \citet{steidel11} reported the exponential scalelength of $r_s$=27.6 kpc, which corresponds to a half-light radius of 46.4 kpc. Thus, we conclude that LAB17139 has a similar size to $z\sim 2.6$ LABs.

In Figure~\ref{fig:lab_size} (right), we also show the line luminosity distribution of known  Ly$\alpha$ blobs at $z$=2--4 \citep{matsuda04,Dey05,Yang10,Erb11,Badescu17}. LAB17139 lies at a relatively high luminosity regime. The size distribution of LABs is more difficult to characterize because measured isophotal size of an LAB is determined by the combination of intrinsic source brightness, redshift, and imaging sensitivity. For example, given everything equal, the same source can have larger isophotal size as the imaging depth increases. In order to construct the intrinsic size distribution of Ly$\alpha$ nebulae, image simulations such as the one adopted here are needed to be run on each of the relevant dataset.

\section{Sky Distribution of Galaxies}
\subsection{A significant overdensity of LAEs at $z=3.13$} \label{LAE_overdensity}

\begin{figure*}[ht!]
\epsscale{1.2}
\plotone{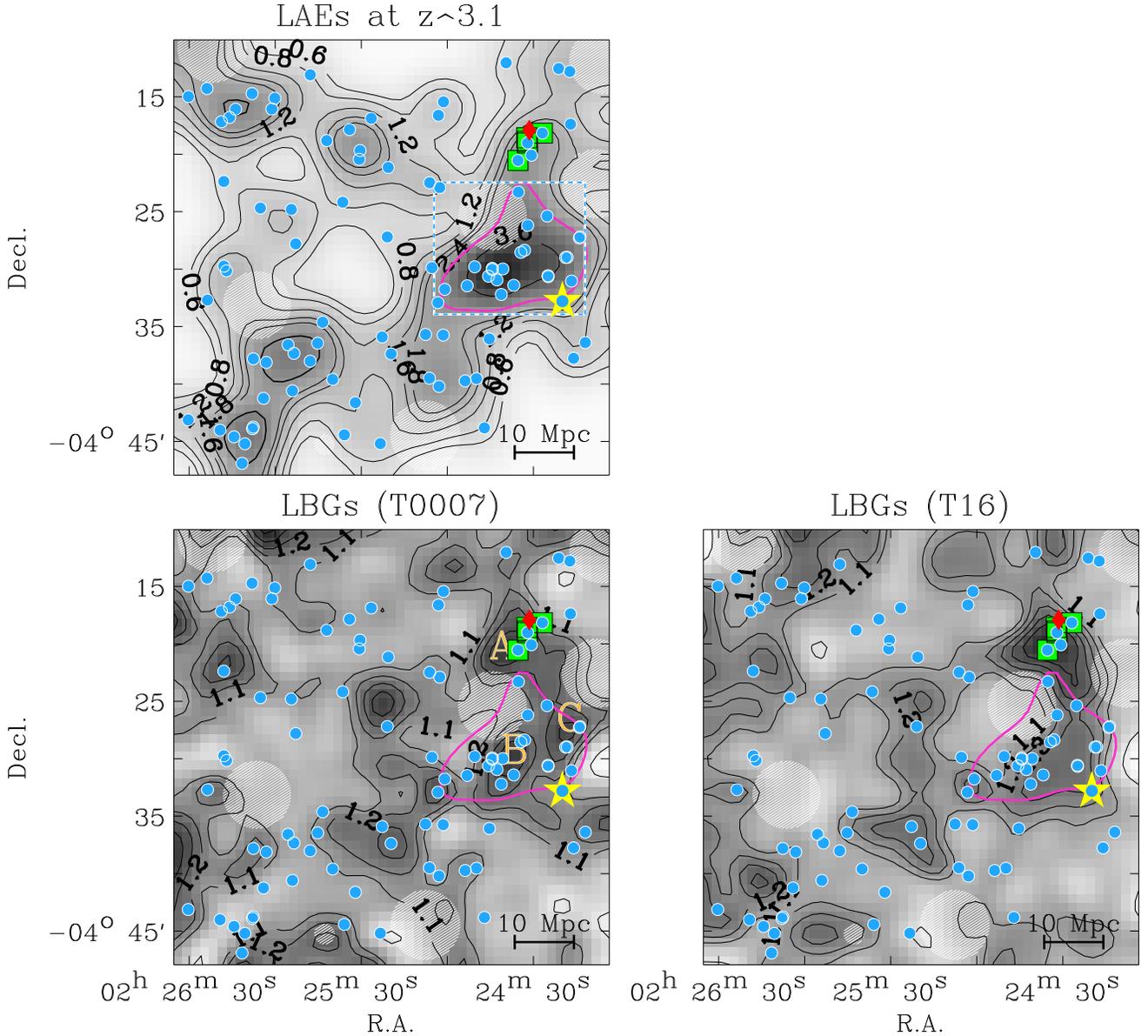}
\caption{
{\it Top left:} a smoothed density map of z=3.13 LAEs. Cyan circles show the LAE candidates while green squares indicate the five spectroscopically confirmed LBGs.  The contours are constructed by smoothing the positions of the LAE candidates with a Gaussian kernel of FWHM=10~Mpc, and the contour labels show  surface density levels relative to the field. The thick solid pink line outlines the LAEs overdensity region. Pixels near bright saturated stars are masked out (hatched circular regions), and do not contribute to the overdensity estimate (see text). The rectangle box is the region where we run K-S test for the LAEs and LBGs distributions. A large Ly$\alpha$ nebula (LAB17139: yellow star) and the brightest $K_S$ band source (red diamond) are also shown. 
{\it Bottom left and Right:} similarly constructed  density maps using LBGs selected from the official CFHTLS catalog  (left) and the \citet{Toshikawa16} catalog (right), respectively. For smoothing, a Gaussian kernel of FWHM=6~Mpc is used (see text for discussion). Three LBG overdensities are labelled as `A', `B', and `C'. Positions of individual LBGs are not shown for clarity; other symbols  are identical to those in the left panel. 
}
\label{fig:sky_distribution}
\end{figure*}

The LAE distribution in the sky appears to be highly inhomogeneous, suggesting that there may be overdense structures. To quantify their spatial distribution, we start by estimating the mean LAE density. After removing the regions near saturated stars (hatched circular regions in Figure~\ref{fig:sky_distribution}), the effective area is 1,156~arcmin$^2$ over which 93 LAEs are distributed. Thus, the LAE surface density is $\bar{\Sigma}$=0.08$\pm$0.01~arcmin$^{-2}$ where the error reflects the Poisson noise. 

To create a LAE density map, we place point sources in the masked regions whose numbers are commensurate with that expected at random locations to avoid producing artificial under-densities. On the positional map containing 93 LAEs and point sources, we apply a Gaussian kernel of a FWHM of 10~Mpc~(5.1\arcmin: $\sigma=4.25$~Mpc). A similar smoothing scale has been used to identify LAE overdensities in the literature \citep[e.g.,][]{Lee14,Badescu17}. The resultant map is shown in the top left panel of  Figure~\ref{fig:sky_distribution} as contour lines and grey shades. The contour line values represent the local surface density relative to the mean value. The positions of individual LAEs are also shown. 

The highest LAE overdensity is located $\sim$5\arcmin\ west of the field center. Twenty one galaxies are enclosed within the purple contour (2.4$\bar{\Sigma}$ iso-density line), within which the effective area is 72.8~arcmin$^2$ (275 Mpc$^2$). 
We choose this region as the LAE overdensity. Scaling from the mean LAE surface density ($0.08$~arcmin$^2$), the expected number of galaxies within this region is $5.8\pm2.4$. Thus, the region contains 3.6 times more galaxies than expected ($\delta_\Sigma \equiv (\Sigma-\bar{\Sigma})/\bar{\Sigma}=2.6\pm 0.8$).

We recompute the mean density after excluding those in the LAE overdensity, and obtain $\bar{\Sigma}$=$0.067\pm0.008$~arcmin$^{-2}$; this estimate is insensitive to inclusion or exclusion of the `D1UD01' region which contains only a few LAEs. The revised overdensity is $\delta_\Sigma= 3.28\pm0.94$. Interestingly, LAB17139 is located at the outskirts of the LAE overdensity (see \S~5.2 for more discussion). 

We  test the robustness of our overdensity estimate by computing the  number of LAEs expected in our survey assuming the field Ly$\alpha$ luminosity functions at $z\sim3.1$ \citep{Gronwall07, Ouchi08}. The expected number of LAEs in a  magnitude bin $[m_k, m_k+\Delta m]$ and redshift bin $[z_j, z_j+\Delta z]$ is:
\begin{equation}
N_{\mathrm{LAE}}(z_j,m_k)=V_jp(m_k)S(z_j)\int_{L_k}\phi(L)dL,
\end{equation}
where $S(z)$ is the normalized redshift selection function, defined from the effective filter transmission $T(\lambda)$ of the $o3$ filter expressed as $S(z_j)\equiv T(1215.67\times(1+z_j))/\mathrm{max}(T)$, $V_j$ is the effective comoving volume, $p(m_k)$ is the completeness limit of the $o3$ image in the magnitude bin, which is derived from our image simulations of point sources.  For our calculation, we use $\Delta z=0.003$ and $\Delta m = 0.1$~mag. The total number of LAEs is $N_\mathrm{LAE}= \sum_j\sum_k N_\mathrm{LAE}(z_j,m_k)$.

The expected number of LAEs in our field is 71$\pm$8 using the \citet{Gronwall07} best-fit parameters and 102$\pm$10 using the \cite{Ouchi08} values. As for the errors, we assume Poisson statistics, which are underestimated as they do not include cosmic variance. The observed number of LAEs in our survey field is consistent with that expected in an average field. Using these values as the field LAE density, the overdensity outlined by the purple contour in Fig~\ref{fig:sky_distribution} is $\delta_\Sigma=2.2 - 3.6$, consistent with our previous estimate.

The significance of the newly discovered LAE overdensity is comparable to those found in known structures in the literature. \citet{Kurk00} reported an LAE overdensity of $\delta_{\Sigma}=3$ around a radio galaxy at $z=2.16$ \citep{Venemans07}. Another radio galaxy at $z=4.1$ is associated with an LAE overdensity of $\delta_{\Sigma}=3.7$ \citep{Venemans02, Venemans07}. The line-of-sight distances probed by these surveys are similar to this study ($\Delta z\approx 0.04-0.05$). \citet{Lee14} reported two structures with similar LAE overdensities, which were later confirmed spectroscopically as protoclusters \citep{Dey16}. 

\subsection{LAE vs LBG distributions}\label{sec:lae_vs_lbg}

If the LAE overdensity we discovered at $z\sim3.13$ represents a genuine protocluster,  the same region is expected to be traced by non-LAEs at the same redshift. Existing observations suggest that LAEs represent a subset of star-forming galaxies likely observed through sightlines with the lowest optical depths \citep{shapley03} and otherwise obey similar scaling relations as UV color-selected star-forming galaxies \citep[e.g.,][]{Lee14,Shi19}. However, some LAEs appear to have lower metallicities, higher ionization parameters \citep{finkelstein11,nakajima13,song14}, and less massive with younger ages \citep[e.g.,][]{Gawiser07,guaita11,Hathi15}  than non-LAEs.

Isolating non-LAEs at the same redshift is a formidable task. In principle, similar to the LAE selection, one can look for narrow-band `deficit' sources to find galaxies with strong Ly$\alpha$ absorption \citep[e.g.][]{Steidel00}. However, the depth of our imaging data is inadequate for this method to be effective. Alternatively, one can use the surface density of LBGs as a proxy to search for  high overdensity regions. Using the Millennium simulations, \citet{Chiang13} demonstrated that the progenitors of the most massive galaxy clusters reside in regions of elevated densities even at the redshift smoothing scale of $\Delta z \lesssim 0.2-0.3$. Observationally,  several confirmed protoclusters are discovered initially as LBG overdense regions \citep[e.g.,][]{Lee14,Toshikawa16}.

To investigate the possibility of LBG overdensities, we use two LBG samples, namely, our fiducial LBG sample selected from the T0007 catalog, and the T16 catalog constructed by \citet{Toshikawa16}.  The number of LBGs in these catalogs are slightly different: 6,913 and 7,793, respectively, which mainly reflects the differences in  their source detection setting as discussed in Section~\ref{subsec:lbg}.

Similar to the LAE density map, we smooth the positions of each LBG using a Gaussian kernel.  In determining the size of a smoothing kernel, two factors need to be taken into consideration:  the source surface density and the volume within which a galaxy overdensity is  enclosed. For instance, using a kernel size smaller than the typical distance between  two nearest neighbors is undesired as most `overdensities' will consist of a single galaxy. On the other hand, using too large a kernel size effectively averages out cosmic volumes that are much greater than a typical size of a galaxy overdensity, thereby washing away the very signal one is searching for. 

In the case of LBGs, the source density is sufficiently high (and the distance to the nearest neighbor small) that the cosmic volume consideration becomes the main determinant of the kernel size.  \citet{Toshikawa16} used a tophat filter with a diameter 1.5~Mpc (physical) in their search of $z\sim3-6$ LBG overdensities. The size was justified as a typical angular size enclosing protoclusters in cosmological simulations \citep{Chiang13}.  At $z=3.13$, this corresponds to 6.2~Mpc. 

In the bottom left and right panels of Figure~\ref{fig:sky_distribution}, we show the resultant LBG maps using a smoothing FWHM of 6~Mpc.  Grey scales and contour lines indicate the density fluctuations together with the positions of the LAEs (cyan circles), and five spectroscopic sources at $z=3.13$ (green squares). 

Several overdensities are present but each with a much lower significance than our LAE overdensity. This is not surprising considering the line of sight distances sampled by them. Our image simulation suggests that the FWHM in the redshift selection function, at $r\approx 24.5$, is $\Delta z=0.7$  corresponding to 666~Mpc. Using the $o3$ filter FWHM, the LAE redshift range is $z$=3.109--3.155, spanning just 44~Mpc in the line-of-sight distance, more than an order of magnitude smaller than that of the LBGs. The large $\Delta z$ of the LBGs can also result in artificial overdensities because of chance alignments along the line of sight. It is easy to understand that, even in a sightline of a massive protocluster, LBGs with no physical association with the structure will outnumber those in it. 

In both LBG density maps (T0007 and T16 LBG samples), smaller overdensities as well as underdense regions spanning $\gtrsim 20$~Mpc are found in identical locations, and the largest and most significant overdensity structures are found at the western end of the field. In the T0007 map, the region consists of three adjacent overdensities labelled as `A', `B', and `C' in Figure~\ref{fig:sky_distribution}.  In the T16 map, the overdensity A is noticeably more pronounced while the B and C overdensities, which are merged into a single overdensity, appear less significant. 

The LAE overdensity largely coincides with the `B+C' region, and stretches toward the `A' region where a concentration of four LAEs lies. It is intriguing that in the `A' region, which \citet{Toshikawa16} found to be the most significant LBG overdensity, relatively few LAEs are found. Assuming that all LAE candidates lie at $z=3.13$, there are a total of just six galaxies in the `A' region (including the spectroscopic sources that barely escape the LAE selection). In comparison, the LAE overdensity contains 21 LAEs. Extending the overdensity region slightly would include additional two LAEs and one Ly$\alpha$ nebula (Section \ref{sec:lab}). 

Comparing the LAE and LBG maps, it is evident that their sky distributions are disparate. Other than the main LAE overdensity, none of the LAE density peaks coincides with the LBG overdensities. This likely suggests that there is only a single large-scale structure that exists at $z=3.13$, and smaller LAE overdensities are a product of Poisson fluctuations, or alternatively, belong to much less significant cosmic structures than the one which the main LAE overdensity inhabits. 

Next, we make a quantitative comparison of the LBG and LAE distributions in the general LAE overdensity region. To this end, we perform a series of two-dimensional Kolmogorov-Smirnov tests \citep{Peacock83,Fasano87}. We define a rectangular region enclosing the LAE overdensity as outlined in the left panel of Figure~\ref{fig:sky_distribution} whose area is 115~arcmin$^2$.  Running the 2D K-S test in the LBG and LAE distributions yields the $p$ value of 0.28 (0.20) using the fiducial (T16) LBG samples. The large $p$ value indicates the similarity of the two distributions.

As the 2D K-S test is less reliable than the one-dimensional test, we  perform a control test to interpret the $p$ values. First, we create two random samples that are uniformly distributed in the rectangular region, each matching the number of LAEs and LBGs in our samples, and calculate the corresponding $p$ value. The process is repeated 1,000 times and the $p$ value is recorded each time. We obtain a median (mean) $p$ value of 0.23 (0.27). Second, we assume that the underlying distribution is a two-dimensional Gaussian function with $\sigma$=5\arcmin\ centered at the middle of the rectangle, and repeat the test, obtaining similarly large $p$ values (median and mean value of 0.27 and 0.31). 

Finally, we  test the similarity of the two galaxy samples in the entire field by moving the rectangle to random locations. Whenever a masked region falls within the subfield, we randomly populate the area with the expected number of point sources therein before performing the test. The median (mean) $p$-value is 4.5$\times 10^{-15}$   ($9\times 10^{-4}$). These tests give strong support to the possibility that the cosmic structure traced by the LAE overdensity is also well populated by LBGs at a level not observed in other parts of the survey field.

All in all, our analyses strongly suggest a presence of a significantly overdense cosmic structure, which includes $21$ LAEs, 5 spectroscopically confirmed $z=3.13$ LBGs, and one luminous Ly$\alpha$ nebula. A large number of LBGs exist in the general region although, without spectroscopy, it is difficult to know how many of them truly belong to the structure. A segregation of the highest LAE and LBG overdensity is also curious. We discuss possible implications of our results in Section~\ref{sec:speculation}.

\section{Discussion}

\subsection{Descendant mass of the protocluster} \label{mass_estimate}

\subsubsection{The present-day mass of the LAE overdensity}
Given that the $o3$ filter samples $\approx$44~Mpc in the line-of-sight direction, a surface overdensity computed based on the angular distribution of galaxies should scale closely with a given intrinsic galaxy overdensity with a minimal contamination from fore- and background interlopers.  In this section, we estimate the true galaxy overdensities, and infer their descendant (present-day) masses.  

Based on the Millennium Runs \citep{millennium}, \citet{Chiang13} calibrated the relationship between  galaxy overdensity ($\delta_g$) and  present-day mass $M_{z=0}$ at a given redshift. Galaxy overdensity is measured in a (15~Mpc)$^3$  volume ($\delta_{g,15}$ hereafter) using the galaxies whose host halos have the bias value of $b\approx2$. This value is comparable to that typically measured for LAEs \citep{Gawiser07,guaita10, Lee14}, which lie at similar redshift and are of comparable line luminosities to those in our sample.  Thus, it is safe for us to apply the Chiang et al. calibration without further corrections. 

The transverse area enclosing the LAE overdensity is 275~Mpc$^2$, reasonably close to that of (15~Mpc)$^2$ used by \citet{Chiang13}. However, the line-of-sight distance sampled by the $o3$ filter is $\sim$3 times larger than their sampled volume. Generally, averaging over a larger volume reduces the significance of the overdensity. We correct this effect using their Figure~13 where they show how, for a fixed $\delta_{g,15}$, measured (surface) overdensity  drops with increasing redshift uncertainty $\Delta z$. Correcting the measured overdensity (\S~\ref{LAE_overdensity}) accordingly results in $\delta_{g,15} = 5.5\pm 1.6$. Inferred from Figure 10 of \cite{Chiang13}, the corresponding descendant mass at $z=0$ is $M_{\rm tot}\approx (0.6-1.3)\times10^{15}  M_{\sun}$. The estimated overdensity well exceeds the value $\delta_{g,15}=3.14$, above which there is $>$80\% confidence that it will evolve into a galaxy cluster by $z=0$. These considerations lend confidence that the newly identified LAE overdensity is a genuine massive protocluster. 

Alternatively, a more empirical method may be employed similar to that taken by \citet{Steidel98}. If all mass enclosed within the overdensity will be gravitationally bound and virialized by $z=0$,  the total mass can be expressed as:
\begin{equation}
M_{z=0}=(1+\delta_m)\langle \rho\rangle V_{\rm true}
\label{eq:mass}
\end{equation}
where $\langle \rho\rangle $ is the mean density of the universe, and $V_{\rm true}$ is the true volume of the overdensity. The matter overdensity $\delta_m$ is related to the galaxy overdensity through a bias parameter. The bias parameter $b$ can be described as $1+b\delta_m=C(1+\delta_g)$ where $C$ represents the correction factor for the effect of  redshift-space distortion due to peculiar velocities. The true volume $V_{\rm true}$ is underestimated by the same  factor as $V_{\rm true} = V_{\rm obs}/C$. In the simplest case of spherical collapse, it is expressed as:
\begin{equation}
\label{eq:c}
C(\delta_m,z)=1+\Omega_m^{4/7}(z)[1-(1+\delta_m)^{1/3}].
\end{equation}
Equation~\ref{eq:c} and the equation relating the matter and galaxy overdensity can be evaluated iteratively to determine $C$ and $\delta_m$. The observed overdensity is $\delta_g=3.3\pm 0.9$ and the estimated survey volume of  $V_{\rm obs}=1.21\times10^4$~Mpc$^3$. The bias value is assumed to be $b\approx 2$ \citep{Gawiser07,guaita10,Lee14}. We obtain the matter overdensity in the range of $\delta_m=0.8-1.3$; thus the total mass enclosed in this overdensity is $M_{z=0}\approx (1.0-1.5) \times 10^{15}~M_{\sun}$, in good agreement with the simulation-based estimate. We conclude the LAE overdensity will evolve into a Coma-like cluster by the present-day epoch \citep[see, e.g., ][]{kubo07}.

\subsubsection{The structure and descendant mass of the LBG overdensity}

As discussed in Section~\ref{sec:lae_vs_lbg}, there appear to be multiple LBG overdensities; these regions are marked as `A', `B', and `C' in Figure~\ref{fig:sky_distribution}. The `B' and `C' overdensities are merged into one in the T16 catalog, and lie in a region largely overlapping with the LAE overdensity, hinting at their physical association. On the other hand, the `A' overdensity -- $\approx 15$~Mpc away from the LAE overdensity -- may be a separate system and is largely devoid of LAEs therein. As such, we consider the `B+C' and `A' as two separate structures and evaluate their significance. 

The LBG color criteria (\S~\ref{subsec:lbg}) typically result in a relatively wide redshift selection, $\Delta z = 0.4-0.6$. The redshift range of spectroscopic sources yields the median redshift of $z=3.2$ with the standard deviation $0.6$, in a reasonable agreement with the  FWHM $\Delta z$ estimated from our photometric simulations (Section~\ref{sec:lae_vs_lbg}). This very wide $\Delta z$ makes it challenging to directly use the \citet{Chiang13} calibration. Instead, we use an alternative method described in \citet{Shi19} to estimate the intrinsic galaxy overdensity as follows. 

We create a mock field, which is of the same size as our survey field and contains a single protocluster with a galaxy overdensity $\delta_g$. The protocluster overdensity is assumed to extend a transverse size of the LBG overdensity (`A') and 15~Mpc in the line-of-sight distance. We divide the redshift range $z=[2.7,3.4]$ into 35 bins each with $\Delta z = 0.02$ ($\sim$19~Mpc). The number of galaxies belonging to the protocluster is then expressed as $N_{\rm proto} = (1+\delta_g) N_{\rm all}/(35+\delta_g)$ where $N_{\rm all}$ is the total number of LBGs in the field. We populate the remainder ($N_{\rm all}-N_{\rm proto}$) at random in the redshift and angular space. As for the protocluster galaxies, they are also  randomly distributed but are confined within the overdense region. Based on the galaxy positions, we construct the surface density map in the identical manner to the real data, and estimate the mean overdensity within the protocluster region. We repeat the procedure 10,000 times while varying the intrinsic overdensity $\delta_g$ in the range of 1--30, and obtain a relationship between the observed surface density and the intrinsic overdensity.

For the level of observed overdensity for the `A' and `B+C' regions, we choose the 1.3$\bar{\Sigma}$ iso-density contour based on our fiducial LBG catalog; the transverse area of these regions are 36 and 27~arcmin$^2$, respectively; the `B' and `C' contours are disjoint and we simply add the enclosed regions. Using our simulation as described above, the intrinsic overdensities of the `A' and `B+C' regions are $\delta_g$ of $(18.4-28.4)$ and $(14.4-23.0)$, respectively. We assume that their galaxy bias is $b_{\rm LBG} \approx 2.6$, i.e., slightly higher than that of the LAEs \citep{Bielby13,Cucciati14}. Using Equation~\ref{eq:mass} again, we obtain the total masses of these structures $(0.6-1.0)\times10^{15} M_{\sun}$ and $(0.4-0.6)\times10^{15} M_{\sun}$, respectively. Increasing the bias value to $b_g=3$ would decrease the mass by 13\%; decreasing it to a value similar to the LAE bias would have the opposite effect on the mass. Generally, our mass estimate is relatively insensitive to a specific choice of isodensity value. This is because lowering the density contrast tends to increase the effective area at a lower overall density enhancement, while raising it has the opposite effect. 

We repeat our mass estimates using the T16 catalog which yields slightly different levels and angular extent of the overdensities. The resultant masses for the two structures are $(0.8-1.1)\times10^{15} M_{\sun}$ and  $(0.4-0.6)\times10^{15} M_{\sun}$. The `A' structure has $\approx 20$\% larger mass, reflecting its more pronounced density contrast in the T16 catalog; the estimate for the `B+C' region is consistent with the earlier estimate. 

Having estimated the LBG-traced total (descendant) masses of both regions, it is worth contrasting them with the  inferred values had we used the LAEs as tracers instead. In the `A' region, only three LAEs exist resulting in an insignificant overdensity $\delta_g=0.6\pm0.3$, and $M_{\rm tot} \sim (2-4)\times 10^{14}M_\odot$, much smaller than the LBG-inferred values. For this calculation, we assume the surface area (36~arcmin$^2$) defined by the LBG distribution, which is clearly much larger than the ill-defined area which the three LAEs populate. Thus, the resultant mass should be regarded as an upper limit.  
As for the `B+C' region, the disagreement of the inferred masses is less severe: we prviously obtained $\sim (0.6-1.3)\times 10^{15}M_\odot$ and $(0.4-0.6)\times 10^{15}M_\odot$ for the LAE- and LBG-based estimates, respectively. If we were to use the surface area of the `B+C' region defined by the LBGs, the former becomes $\sim(0.6-0.9)\times10^{15} M_{\sun}$, further alleviating the tension. Nevertheless, it is obvious that these galaxy types do not yield consistent mass estimates. In Section~\ref{sec:speculation}, we discuss several physical scenarios which may be responsible for this disagreement. 

\subsection{On the possible  configuration of the structures and their constituents}\label{sec:speculation}

The observational data presented in this work paint an incomplete picture leaving several unanswered questions. First, the spatial segregation between the LAE- and LBG-traced structure is puzzling because the spectroscopically confirmed sources in the latter lie at the same redshift as the former.  If the galaxies in the LBG overdensity trace a single structure, the implication would be that the structure `A' genuinely lacks Ly$\alpha$-emitting galaxies whereas the structure `B+C'  is populated by both LAEs and LBGs (and one large Ly$\alpha$ nebula). 
 
The expected comoving size of galaxy clusters observed at $z\sim3$ ranges in 15--20~Mpc \citep{Chiang13}, in agreement with recent estimates of several confirmed protoclusters \citep{Dey16,Badescu17}. In comparison, the projected end-to-end size of the combined structure, at $\approx 30$~Mpc, is simply too large, suggesting that the two are two separate structures. If they lie at the same redshift (i.e., the projected distance is close to the true separation), the dynamical timescale ($\tau\sim \sqrt{ R^3/GM}$) is in the order the Hubble time. 

If `A' and `B+C' represent two separate systems,  we speculate several physical scenarios consistent with the current observational constraints. First, we may be witnessing galaxy assembly bias: a baryonic response to the well-known {\it halo assembly bias}. The latter generally refers to the fact that the spatial distribution of dark matter halos  depends not only on mass but also on other properties such as concentration parameter, spin, large-scale environment, and halo formation time \citep[e.g.,][]{gao05,wechsler06,li08,zentner14}. In the present case, a given halo's environment and formation time are particularly relevant considering that clusters are expected to be the sites of the earliest star formation in the densest environment. 

\citet{zehavi18} examined the importance of the halo formation time and its large-scale environment in determining the halo mass ($M_h$) to stellar mass ($M_{\rm star}$) and the halo mass to the clustering strength scaling relations. They found that,  while controlling for $M_h$, halos with an earlier formation time tend to host galaxies with larger $M_{\rm star}$, have fewer satellites, and are more strongly clustered in space compared to those that formed at a later time. Similar dependence was found for the halos' large-scale environment (measured as dark matter density smoothed in a $5 h^{-1}$~Mpc scale where $h=0.7$) where a higher density and earlier formation time have similar effects on galaxies' properties. 

In this context, we speculate that the `B+C' structure  may have formed more recently than the `A', and thus is traced by numerous young and low-mass galaxies, many of which are observed as LAEs \citep[see, e.g.,][]{guaita10, guaita11}. In comparison, as an older and more settled system, the `A' protocluster has had more time to accrete  surrounding matter and to merge with lower-mass satellite halos, and thus is expected to have more evolved (i.e., more stars and dust) star-forming galaxies that are observed as LBGs, while having fewer low-mass systems such as LAEs. \citet{zehavi18} also found a strong variation in their clustering amplitude at scales 5--10~Mpc (see their Figure~10). One implication may be that  it is not appropriate to use a galaxy bias representative of the field galaxy population of the same type (as we have done in Section~\ref{mass_estimate}) in estimating the total descendant mass of a structure that likely formed the earliest.  
 
A slight variant of the above hypothesis is that, while the two have similar present-day masses, `A' is simply a more massive and relaxed halo  at the time of observation than `B+C', which is an aggregate of two or more of smaller halos. Similar environment-dependent processes are expected to those in the assembly bias scenario. 

In these conjectures, it follows that both  LAE- or LBG-based protocluster searches would be sensitive to different evolutionary stages (or ages) of cluster formation in which the former (latter) method favors younger (older) structures. Similarly, the presence of galaxies with old stellar populations should be predominantly found in the LBG-selected structures but not in LAE-traced ones. 
While several known structures  support this picture \citep{steidel05, wang16, Shi19}, only a handful of protocluster systems have been characterized using multiple galaxy tracers \citep[including LAEs: e.g.,][]{Steidel00,kurk2000,kurk2004} in a similar manner to the present work, making it difficult to evaluate the validity of such a hypothesis. A rigorous test would require a well-controlled statistical approach in which large samples of LBG- and LAE-selected overdensities are identified independently and compared for the level of their cohabitation.  The same test can also inform us about how baryonic physics can impact the manner in which galaxies trace the underlying large scale structure in dense environments \citep[e.g.,][]{orsi2016}. 

The final and most innocuous scenario to explain the curious configuration of `A' and `B+C' is as follows: the spectroscopically confirmed sources embedded in the `A' region may be spatially disjoint from the majority of LBGs therein, and are part of a small group falling in towards the `B+C' structure. The `A' region then could represent just another protocluster with no physical association with the LAE overdensity. It is unlikely, however, because within the `A' region \citet{Toshikawa16} confirmed 30 galaxies between $z=2.73$ and $z=3.56$, and there was only one significant redshift overdensity at $z=3.13$.   More extensive spectroscopy in all of the `A+B+C' regions can elucidate the true configurations of these structures. \\

Finally, we contemplate on the significance of  the Ly$\alpha$ nebula in the context of protocluster formation. As described in  \S~\ref{sec:lab}, our search of the entire field resulted in a single LAB. The fact that it is located at the southwestern end of the LAE overdensity (`B+C') is significant. 

There is mounting evidence that luminous Ly$\alpha$ nebulae are preferentially found in dense environments. \citet{matsuda04} identified 35 Ly$\alpha$ nebulae candidates in an LAE and LBG rich protocluster at $z=3.09$, and reported that the LAEs and LABs trace one another. \citet{Yang10} conducted a systematic search for LABs in four separate fields, each comparable in size to our survey field. The number of LAB they identified in each field ranged in 1--16; they argued that high cosmic variance implies a very large galaxy bias expected for group-sized halos. Small groups of galaxies are observed to be embedded in several luminous blobs \citep[e.g.,][]{Dey05,prescott12,yang11,Yang14} in agreement with \citet{Yang10}'s assessment. 

It is notable that LAB17139 lies at the periphery of the `B+C' overdensity traced by LAEs. Recently, \citet{Badescu17} compiled the LAE/LAB data for five protoclusters at $z=2.3$ and $z=3.1$, and  showed that LABs are preferentially found in the outskirts of each of the LAE overdensities. They speculated that these blobs may be signposts for group-sized halos (harboring galaxy `proto-groups') falling in towards the cluster-sized parent halo traced by LAEs where Ly$\alpha$-lit gas traces the stripped gas from galaxy-galaxy interactions. 

A significant variation of their numbers implies a relatively short timescale for the LAB phenomenon; that combined with their preferred locations at the outskirts  requires a physical explanation involving the proto-cluster environment. The kinematics of protocluster galaxies showing relatively low velocity dispersions\footnote{ \citet{matsuda05} reported a much larger velocity dispersion of $\sim$1100 km s$^{-1}$ for the SSA22a protocluster at $z=3.09$; however, the spectroscopic LAEs have at least three separate groups. We estimate that the velocity dispersion of each group does not exceed 500 km s${-1}$ \citep[see][for detail]{Dey16}. } and in multiple groupings \citep[$\lesssim$ 400 km s$^{-1}$:][]{Dey16} indicate that the structure is far from virialization. 

If a galaxy overdensity is a superposition of multiple overdensities in physical proximity, LAB's preferred location at their outskirts may signify their first group-group interactions enabling a host of galaxy-galaxy interactions which in turn bring about starbursts, AGN, and stripped gas lighting up an extended region surrounding these galaxies.

\subsection{The physical properties of LAEs and their environmental dependence}

We investigate whether local environment influences the properties of the LAEs. To this end, we define two LAE subsamples according to their measured galaxy surface density. The `overdensity' sample includes 21 LAEs within the purple contour shown in Figure \ref{fig:sky_distribution} as well as three of the \citet{Toshikawa16} galaxies that we recover as LAEs. The remaining 69 LAEs belong to the `field' sample.  

Apart from the line luminosities and EWs (\S~\ref{photometry}), we also convert the measured UV continuum slope, $\beta$, to the extinction parameter E($B-V$) assuming the  dust reddening law of local starburst galaxies \citep{calzetti00}. 
For the sources with relatively robust  $\beta$ measurements ($\Delta \beta < 0.9$), we also derive dust-corrected SFRs by correcting the continuum luminosity accordingly using the \citet{kennicutt} calibration. In the overdensity and field sample, 21 (88\%) and 42 (61\%) LAEs  have the SFR estimates. The difference stems from the fact that the former sample is on average more UV-luminous (see later). 
 However, our SFR estimates are only approximate given a relatively large uncertainty in the measured UV slopes; increasing (decreasing) $\beta$ value by $\Delta \beta =0.4$ (which is well within a typical uncertainty) would lead to a 41\% increase (58\% decrease) in the SFR estimate.

\begin{deluxetable*}{ccccccc}
\tablecaption{Key physical properties of LAEs in different environments}
\tablehead{
                   & $N$ & \colhead{W$_{0,{\rm Ly}\alpha}$} & \colhead{log($L_{{\rm UV},{\rm obs}}$)} & \colhead{log($L_{{\rm Ly}\alpha,{\rm obs}}$)} & \colhead{E($B-V$)} & \colhead{SFR$_{\rm UV,cor}$}\\
\colhead{} &\colhead{} & \colhead{[\AA]} &\colhead{[erg~s$^{-1}$ Hz$^{-1}$]} & \colhead{[erg~s$^{-1}$]} & \colhead{} & \colhead{[\msun yr$^{-1}$]} }
\startdata
 & & & All Galaxies \\
\hline
Overdensity & 24 & $75~(57) \pm 16$  & $28.05~(28.08) \pm 0.08$   & $42.57~(42.53) \pm 0.05$                                                    &- &                               -        \\
Field        & 69  & $104~(53) \pm 16$      & $27.81~(27.84) \pm 0.06$   & $42.45~(42.38) \pm 0.04$                                                    & -  &                -                          \\
All       &93     & $96~(55) \pm 13$         & $27.87~(27.93) \pm 0.05$   & $42.52~(42.42) \pm 0.03$                                                    & -  &                 -                    \\ 
\hline
& & & Galaxies with $\beta$ Measurements & \\ 
\hline
Overdensity & 21 & $52~(49) \pm 5$  & $28.20~(28.24) \pm 0.07$   & $42.67~(42.65) \pm 0.05$                                                    & $0.11~(0.06) \pm 0.03$ & $15~(6) \pm 6$     \\
Field  & 42        & $62~(46) \pm 9$ & $28.00~(27.99) \pm 0.06$   & $42.48~(42.38) \pm 0.05$                                                    & $0.13~(0.14) \pm 0.02$   & $10~(5) \pm 2$   \\
All      & 63      & $59~(47) \pm 6$  & $28.06~(28.05) \pm 0.05$  & $42.52~(42.42) \pm 0.04$                                                    & $0.13~(0.10) \pm 0.02$   & $11~(5) \pm 2$    
\enddata
\tablecomments{The values represent means of key physical properties (medians in the brackets) with uncertainties for each sample.}
\label{ksresults}
\end{deluxetable*}

\begin{figure*}
\epsscale{1.1}
\plotone{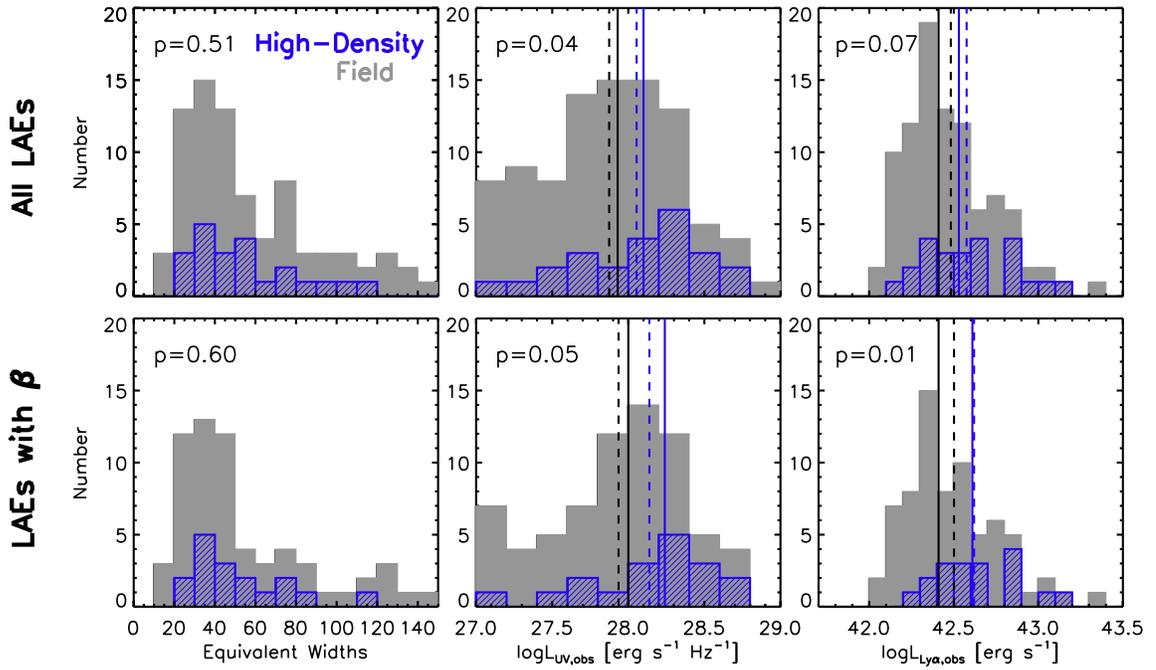}
\caption{
Histograms of rest-frame equivalent widths (left), observed UV luminosity (middle), and observed Ly$\alpha$ luminosity (right) of two subsamples. Blue hatched and grey histograms represent high-density LAEs and field LAEs, respectively. The median $p$ values obtained from the Kolmogorov-Smirnov test are shown in each panel. The vertical solid lines are the mean values while the dashed lines are the median values.
   }
\label{hist}
\end{figure*}

The mean properties of each subsample are listed in Table~\ref{ksresults} with the errors corresponding to the standard deviation of the mean, and the overall distributions of these parameters are illustrated in Figure~\ref{hist}. In both, we show our results for the full sample containing 93 LAEs (top), and for the 63 LAEs with reliable SFR estimates (bottom). We find that our conclusions do not change depending on which sample we consider. 

In terms of both line and continuum luminosities, we find a possible enhancement  for the LAEs in the overdense regions compared to those in the field.  The enhancement in UV luminosity is 74$\pm$32\% if we compare all LAEs in both samples, and 58$\pm$22\% if only the LAEs with robust $\beta$ measurements are considered. As for  Ly$\alpha$ line luminosity, the enhancement relative to the field is 32$\pm$15\% and 55$\pm$18\%  for all LAEs and those with $\beta$ measurements, respectively. The median EW and E($B-V$) values are comparable in both samples.

To assess the similarity of the overall distribution of the physical quantities between the two samples, we perform the one-dimensional K-S test. The $p$ values obtained for each distribution are indicated in Figure~\ref{hist}. The values obtained  for the Ly$\alpha$ and UV luminosity distributions lie around $p\sim 0.05$ corresponding to a $2\sigma$ in the confidence level. As for the EWs and UV slopes $\beta$, the distributions are statistically indistinguishable for the two environmental bins. 

\begin{figure*}
\epsscale{1.2}
\plotone{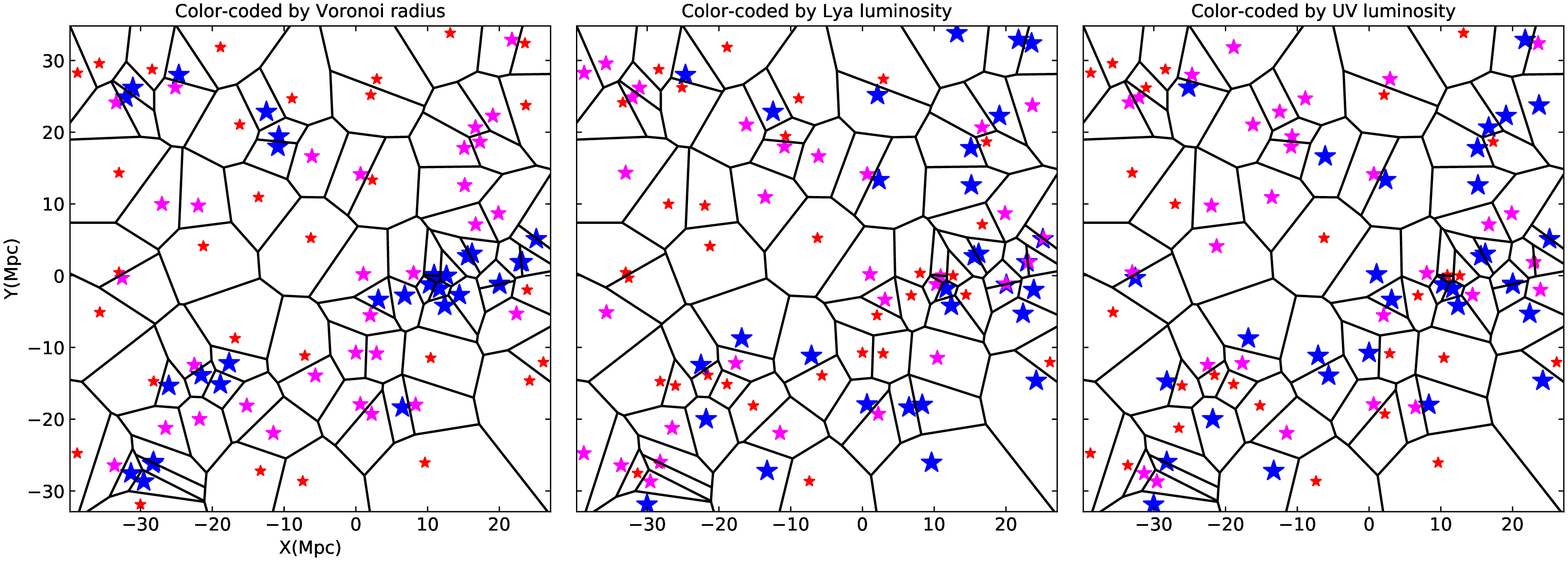}
\caption{
Voronoi tessellated maps of the LAE positions. The LAE sample is divided by Voronoi radius (left), Ly$\alpha$ luminosity (middle), and UV luminosity (right) of each LAE. In all panels, blue, pink and red colors are used for the top, middle, and bottom third, corresponding to $r_V \leq 3.2$, $3.2 < r_V \leq 4.85$, and $r_V > 4.85$ in Voronoi radius,  log($L_{{\rm Ly}\alpha}$) $>$ 42.60, 42.38 $<$ log($L_{{\rm Ly}\alpha}$) $\leq$ 42.60, and log($L_{{\rm Ly}\alpha}$) $\leq$ 42.38 in line luminosity, and log($L_{\rm UV}$) $>$ 28.06, 27.68 $<$ log($L_{\rm UV}$) $\leq$ 28.06, and log($L_{\rm UV}$) $\leq$ 27.68 in UV luminosity. The correlation between these parameters is evident as a large fraction of UV-/Ly$\alpha$-luminous galaxies are found in the LAE overdensity region.
}
\label{voronoi}
\end{figure*}

The same trend is visualized in Figure~\ref{voronoi}. In the left panel, we show the LAE positions overlaid with a two-dimensional Voronoi tessellated map of the whole field \citep{marinoni02,cooper05}. Each LAE is embedded in a Voronoi polygon with an area $A_V$, and its 2D density scales inversely with the radius of the equivalent circular region  defined as $r_V\equiv\sqrt{A_V/\pi}$. The map is color-coded by the 2D density, with the size of each star increases  with increasing density. The LAE overdensity clearly stands out  as a region with the highest concentration of blue stars.

In the other two panels of Figure~\ref{voronoi}, we show the same tessellated LAE map but the LAEs are color-coded by Ly$\alpha$ (middle) and UV luminosity (right), respectively. A large fraction of blue stars representing a top third populate the combined region of the LBG and LAE overdensities. The trend is particularly evident for the case of continuum luminosity (right panel). Of the total 30 blue stars, 17 (57\%) reside within the LAE overdensity region. No radial dependence is found for the luminosity enhancement within the group although our sample may be too small to discern any trend.

The overall correlation between the LAE density and its luminosity, and the level of enhancement are consistent with the similar trends we reported in \citet{Dey16} for the constituents of another  protocluster at $z=3.78$. The present work takes a step further by examining the UV and line luminosity of the same galaxies, which was not possible previously due to the relatively shallow depth of the broad-band data.

The higher UV- and Ly$\alpha$ mean luminosities observed for protocluster LAEs are curious and cannot be fully explained by the level of overdensity. If the protocluster LAEs obey the same UV or Ly$\alpha$ luminosity function as measured in the field but are simply scaled up by a factor of $(1+\delta_g)$, the expected mean or median values would be identical to those in the field. Observationally, the trend can be due to either the lack of low-luminosity (and low-mass) galaxies or the excess of high-luminosity LAEs in overdense environments relative to the field. 

The lack of low-luminosity, low-mass galaxies near cluster-sized halos may be caused by a variety of astrophysical processes. \citet{orsi2016} showed that AGN feedback (from a quasar or radio galaxy hosted by the central halo) can alter the clustering and abundance of galaxies inhabiting the satellite halos. They also noted that the spatial distribution of LAEs may be more affected than other galaxy tracers -- such as H$\alpha$ emitters -- due to complex radiative transfer effects. \citet{Cooke14} studied a radio galaxy MRC~2104-242 at $z=2.5$ and reported the lack of low-mass galaxies ($M_{\rm star}\lesssim 10^{10}M_\odot$) in their 7~arcmin$^2$ survey field. It is certainly conceivable that there was a radio galaxy or quasar in the past (which has since then turned off), which had  influenced the formation histories of these galaxies. While these trends  warrant further exploration with upcoming surveys such as DESI, it is far from conclusive at this time given the small sample size and areal coverage. Further, the most likely candidate hosting a powerful AGN is not at the center but at the outskirts of this structure.  Future availability of deep near-infrared imaging would be helpful in closer examination of  the stellar mass distribution  and its radial dependence, which can  be compared with those in galaxy simulations.

Alternatively, our results can be interpreted as a mild but widespread enhancement of star formation in the protocluster LAEs.
 One possible explanation for the higher luminosity value may be that the luminosity function (and SFR function)  is more `top-heavy' in protocluster environment, producing a larger fraction of UV-luminous galaxies. This may be brought on by faster-growing halos as suggested by \citet[][]{Chiang17}, or by a different star formation efficiency in clusters whereby a galaxy is more luminous at a fixed halo mass (Y.-K. Chiang, in private communication). Alternatively, it is also possible that protocluster LAEs simply have different ages and/or metallicity than elsewhere; however, the overall similarities in observed colors and EWs in the two environmental bins studied here argues agains this possibility.

Our result is seemingly at odds with some of the existing studies which found that galaxies in dense environment largely grow at a similar rate as those in average fields \citep{Lemaux18, Shi19}, perhaps with an exception at the massive end \citep[e.g.,][]{Lemaux14}. However, it is worth noting that these  studies focused on  more UV-luminous, LBG-like galaxies that are, on average, a factor of $\gtrsim 5-10$ more massive than the LAE population studied in this work. To discern a clearer trend and to study how it depends on galaxy's luminosity and stellar mass, and on galaxy types (LBGs, LAEs, etc.), a more comprehensive study is needed.

We speculate a potential implication of our result in the cosmological context. By following the structures identified as cluster-sized dark matter halos at $z=0$ in the Millennium simulations, \citet{Chiang17} estimated that the fractional contribution to the total star formation rate density (SFRD) from galaxies that will end up in clusters increases dramatically with redshift, from only a few percent at $z=0.5-1.0$, to  $\approx$20--30\% at $z=2-4$, and  to nearly 50\% at $z > 8$. This change is mainly driven by large cosmic volumes occupied by protoclusters well before their final coalescence (see their Figure~1) as well as high galaxy overdensities and  the  top-heavy halo mass function therein \citep[][]{Chiang17}. 

If the observed higher luminosity of protocluster LAEs has an astrophysical origin (e.g., a higher efficiency in converting gas into stars) rather than a cosmological one, it would follow that the total contribution to the cosmic SFRD from protoclusters would be even greater than the \citet{Chiang17} estimate. Separating out these effects will be challenging, however, and will require a much larger sample of protoclusters and a better characterization of halo statistics in different environments.

\subsection{Search of progenitors of a brightest cluster galaxy}

Brightest cluster galaxies (BCGs) are the most massive galaxies in galaxy clusters. In the local universe, they are typically elliptical galaxies residing near the cluster center defined by  X-ray emission peak \citep[e.g.,][]{Lin04}. Identification and characterization of their progenitors (`proto-BCGs') at high redshift would illuminate the early stages of their formation. 

At $z\sim 3$, the rest-frame optical/near-IR luminosity ($0.5-1.6\mu$m) tracing the total stellar content is redshifted into the $K_S$ band and beyond. Thus, the most effective search should be based on the photometric properties at  infrared wavelengths. Although the D1 field was imaged in the near-IR $JHK_S$ bands by the WIRCam Deep Survey  \citep{Bielby12}, the newly discovered galaxy overdensities unfortunately lie near the edge of its coverage (see their  Figure~2 for the coverage map). $\approx$20\% of the area enclosing the `A' and 'B+C' structures has no $K_S$ band coverage while an additional 10\% of the area has only partial coverage ($<$50\% of the full exposure 4.7~hr).  

Given this limitation, we caution that any search based on the existing data would be severely limited by the  depth and areal coverage in obtaining a complete census of massive galaxies in this structure.  While a more comprehensive search of massive galaxies in this region based on the {\it Spitzer} IRAC 3.6$\mu$m detection will be presented in the future (J.~Toshikawa et al., in prep, K.~Shi et al., in prep), in this work, we base our proto-BCG search on our existing LBG catalog instead, focusing on UV-luminous galaxies that already have a large stellar content. We require that a given galaxy must have the $r$-band magnitude $r\leq 24$  \citep[roughly corresponding to $\gtrsim 1.6L^*_{{\rm UV}, z\sim3}$:][]{reddy09}. In addition, to further constrain its stellar mass, it should also be well detected in the $K_S$ band catalog. A total of 80 galaxies satisfy these criteria corresponding to a surface density of 0.06$\pm$0.01 arcmin$^{-2}$.  

\begin{figure*}[ht!]
\epsscale{1.1}
\plottwo{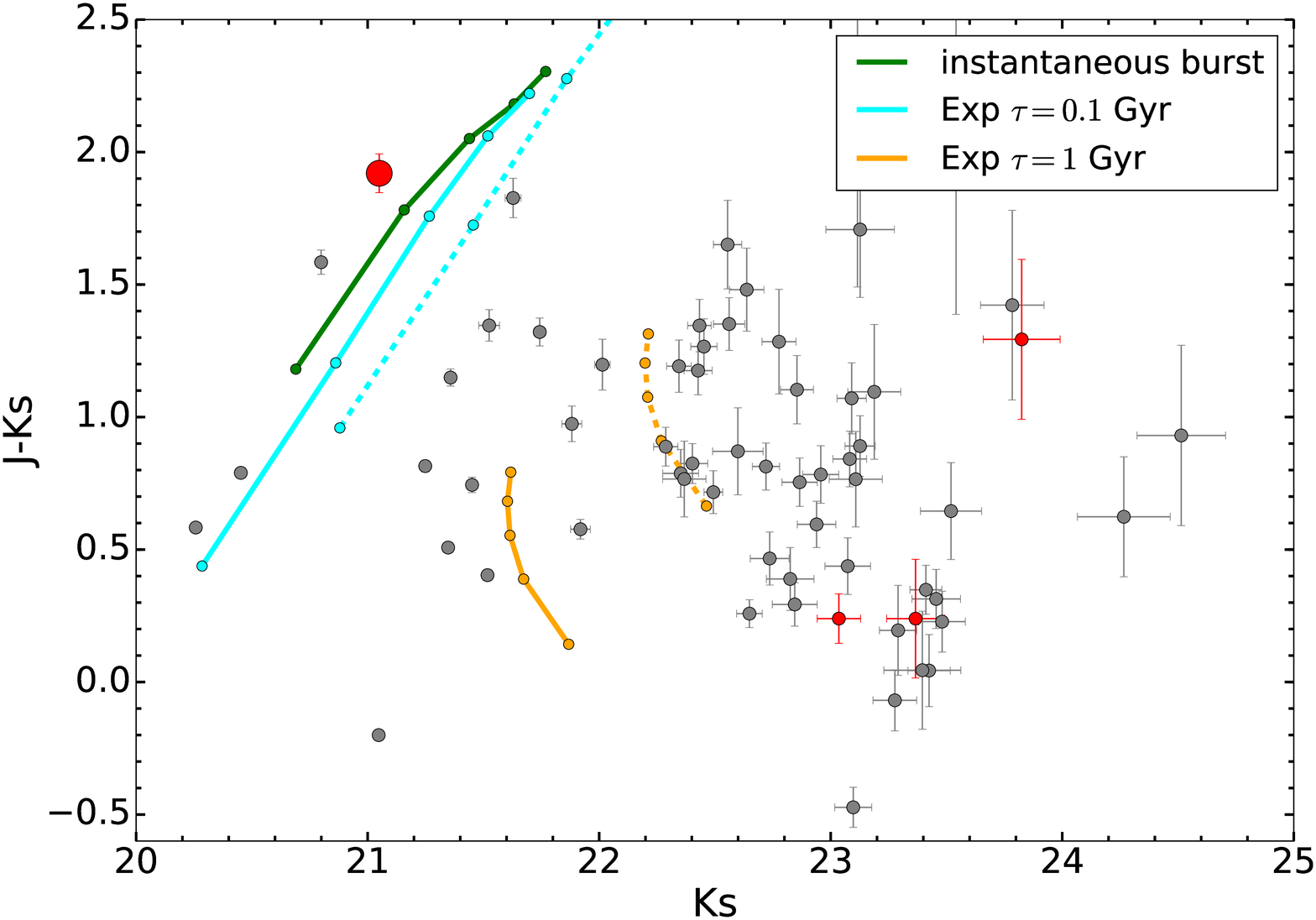}{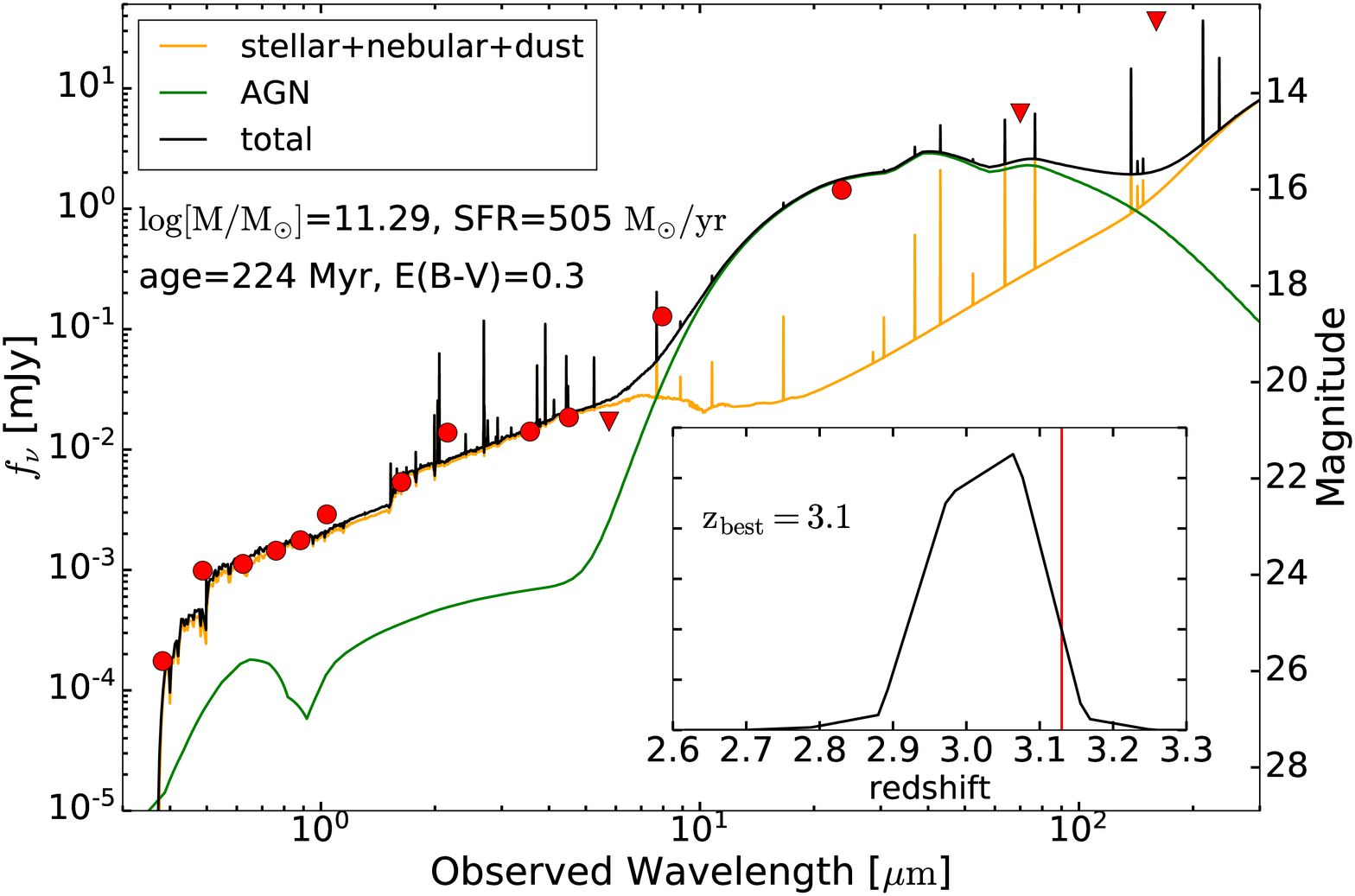}
\caption{
{\it Left:} A color-magnitude diagram of  UV-luminous LBGs with $K_S$ band detection ($r\leq24$: grey circles). The proto-BCG candidate, G411155 (large red circle), is the reddest LBG and also one of the brightest in the $K_S$ band. Three other LBGs near the LAE/LBG overdensities are shown as smaller red circles. Colored lines represent galaxies that formed through an instantaneous burst (green), and with an exponentially declining SF history with $\tau$ values of 100~Myr (cyan) and 1~Gyr (yellow); all are observed at $z=3.1$ with the population age of 0.2~Gyr (bottom) to 1.0~Gyr with a stepsize of 0.2~Gyr. For the exponentially declining models, we also show the color tracks assuming the reddening ${\rm E}(B-V)=0.2$ as dashed lines. 
{\it Right:} The best-fit SED model of G411155 is shown in black together with the photometric measurements (red symbols) and best-fit parameters. The galaxy  (stars+gas+dust) and AGN components are shown in orange and green, respectively. In the inset, we show the photometric redshift probability density. The red vertical line marks $z=3.13$, the redshift of the spectroscopic sources within the LBG overdensity. 
}
\label{fig:bcg}
\end{figure*}

In the left panel of Figure~\ref{fig:bcg}, we show the $J-K_S$ colors vs $K_S$ band magnitudes for all selected sources. The majority have $K_S>22.5$ and relatively blue $J-K_S$ colors. Using the EZGal software \citep{Mancone12}, we also compute the expected color and luminosity evolution assuming several different star formation histories. The \cite{Bruzual03} stellar population synthesis models and local starburst-like dust reddening curve \citep{calzetti00} are adopted for the calculation. The model magnitudes are normalized to a lower redshift ($z=1.8$) $M_{\star}$ cluster galaxy in the $3.6 \micron$ band \citep{Mancone10}, assuming passive evolution from $z\sim3.1$ to $z\sim1.8$. The model tracks represent the time evolution of the galaxy. 

Of the 80 galaxies, only four reside in the combined LBG  and LAE overdensity region. The area covered by the WIRDS data is 93.7 arcmin$^2$ , and thus the expected number therein is $6\pm2$.  The $K_S$-brightest galaxy ($K_S=21.05$), which we dub G411155, is shown as a red circle in the left panel of Figure~\ref{fig:bcg}. Its location is also marked in Figure~\ref{fig:sky_distribution} (red diamond). G411155 is the reddest LBG in the entire field ($J-K_S=1.92$), and would easily meet a typical color selection for distant red galaxies (DRGs) at high redshift \citep[$J-K_S>1.4$:][]{Franx03, Van03}.  G411155 is very bright in the IRAC 8$\mu$m and the MIPS 24$\mu$m bands, having the flux densities of 0.13~mJy and 1.43~mJy, respectively. The remaining three galaxies have relatively modest $K_S$ band brightness  ($K_S=23-24$) and are bluer ($J-K_S<1.4$). None of the five galaxies has a X-ray or radio counterpart \citep{Bondi03, Chiappetti05}.

Using the \cite{Bielby12} and \cite{Lonsdale03} catalogs, we extract the multi-wavelength photometry ($ugrizJHK_S$[3.6][4.5][5.8][8.0][24][70][160]) of G411155 using the  Kron-like total fluxes. We perform the SED fitting with the CIGALE software \citep{Noll09,Boquien18} using both galaxy and AGN templates. Star formation histories are modeled as an exponentially declining function with the characteristic timescale $\tau$ values of 100~Myr to 1~Gyr. AGN models from \cite{Fritz06} are used as templates. 

In Figure \ref{fig:bcg} (right), we show the best-fit SED model together with the photometric measurements. In the inset, we also show the photometric redshift probability density which peaks  at $z\sim 3.1$. The best-fit physical parameters suggest that G411155 has a short star-forming time scale with the luminosity-weighted age of $\approx$200 Myr.  The model fit also suggests a dust-obscured AGN component which dominates the infrared energy budget: 70\% of the total IR luminosity originates from the AGN. The galaxy is already ultramassive at $M_{\rm star}\approx 2\times 10^{11}M_\odot$, and it continues to form stars at a rate ${\rm SFR}\sim 500 M_\odot {\rm yr}^{-1}$! 

We compare our proto-BCG candidate with those found in the literature. \cite{Lemaux14} identified a proto-BCG candidate in a $z\sim3.3$  protocluster. It contains a powerful Type~I AGN (relatively unobscured by dust with broad lines) with a $K_S$ band magnitude of 20.67 ($z-K_S=0.1$) with the estimated stellar mass and age of $\sim8\times10^{10}M_\odot$ and $\sim 300$ Myr. The SFR inferred from the total IR luminosity is $\sim 750~ M_\odot {\rm yr}^{-1}$. Although we do not have a spectrum for G411155, these two proto-BCG candidates have comparable physical properties. 

Near the center of a protocluster at $z=3.09$, \cite{Kubo15,Kubo16} discovered a dense group of massive galaxies  consisting of seven $K_S$ bright ($K_S\sim 22-24$) and red galaxies ($J-K_S>1.1$) with a combined stellar mass of $\approx 6\times10^{11}M_\odot$. They argued that the group is likely in the merger phase which will evolve into a BCG observed in the local universe.  \cite{wang16} reported an overdensity of 11 massive ($\gtrsim 10^{11}M_\odot$) DRGs within a compact core (80~kpc) in another $z=2.51$ structure, and speculated that their findings may signify a rapid buildup of a cluster core. Identifying and studying similar systems in a larger sample of protoclusters will elucidate evolutionary stages of cluster BCGs. 

Finally, G411155 lies very close to the spectroscopic sources at $z=3.13$, in particular four sources two of which are also LAEs, as illustrated in Figure~\ref{LAE_overdensity}. Given its photometric redshift (see inset of Figure~\ref{fig:bcg}), it is possible that these galaxies are members of the same group, which is falling towards the center of its parent halo located at a projected distance $\approx$1~Mpc (physical) away from it. Given its optical brightness, it should be relatively easy to measure its redshift and thereby unambiguously determine its physical association with these sources. 

\section{Summary}

In this paper, we initially set out to investigate a large-scale structure around a significant LBG overdensity in the CFHTLS D1 field. A subset of these galaxies were targeted by \citet{Toshikawa16}, and five  are  confirmed to lie at $z_{\rm spec}=3.13$. At this redshift, Ly$\alpha$ emission is conveniently redshifted into a zero-redshift [O~{\sc iii}] filter, providing a rare opportunity to examine how the same structure is populated by galaxies of different spectral types, thereby evaluating the efficiency of different search techniques for high-redshift protoclusters.  To this end, we have obtained new deep observations using the Mosaic $o3$ filter; by combining the data with the existing broad-band observations, 93 LAE candidates are identified at $z=3.11-3.15$.

The angular distribution of these LAEs is clearly non-uniform, revealing a prominent overdensity   at the western end of the field containing 21 galaxies along with a luminous Ly$\alpha$ nebula. The angular size and level of the  LAE overdensity are consistent with those observed for several confirmed protoclusters. However, our comparison of the LAE and LBG distributions has resulted in a surprising discovery: the LAE-rich region is spatially offset by $\sim$15 Mpc from the LBG-rich region. In the latter, there is a general dearth of LAEs while the LAE overdensity is also populated by LBG candidates. Our findings paint a more complex picture of cluster formation in which the halo assembly bias may play a significant role in determining a dominant type of galaxy constituents therein. Based on our investigations, we conclude the following:\\

\noindent - We report a significant LAE overdensity located 10\arcmin\ south of the five spectroscopic sources at $z=3.13$. The observed surface density therein is higher than that expected in an average field by a factor of $3.3\pm0.9$. The total mass enclosed in the overdensity is estimated to be $M_{\rm tot}\approx (1.0-1.5)\times 10^{15}M_\odot$, implying that the LAE overdensity traces a massive structure that will evolve into a galaxy cluster similar to the present-day Coma. \\

\noindent - We analyze the LBG overdensity based on the existing deep broad-band observations to evaluate its significance and contemplate on its possible relationship with the LAE-traced protocluster. Given the angular extent and the level of overdensity, we conclude that it will also evolve into a Coma-sized galaxy cluster.  \\
 
\noindent - If the spatial segregation of the LAE and LBG-rich structures is interpreted as a manifestation of the halo assembly bias, it follows that different search techniques would be biased accordingly to the formation age of the host halo. Similar selection biases are expected if more massive and relaxed halos preferentially host more evolved galaxies such as LBGs. With multiple upcoming wide-field surveys will be targeting both types of galaxies (e.g., Hobby-Eberly telescope dark energy experiment, Large synoptic survey telescope),  testing this hypothesis will be within reach in the next decade. Such studies will lead us to deeper understanding of early stages of galaxy formation in dense cluster environments, and help us optimize search techniques to reliably identify and study progenitors of massive galaxy clusters. \\

\noindent - We find tentative evidence that the median SFR is higher for Ly$\alpha$-emitting galaxies in protocluster environment. When our LAE candidates are split accordingly to their 2D environment, the LAEs residing in the overdensity consistently have larger  Ly$\alpha$ and UV luminosities -- by $\sim 40$\% and $\sim70$\%, respectively -- than  the rest, in agreement with our previous study based on another protocluster \citep{Dey16}. The enhancement appears to be widespread within the overdensity region with no clear radial dependence. The difference cannot be  explained by the galaxy overdensity alone, and may require either a top-heavy mass function or a higher star formation efficiency for protocluster halos. However, we cannot rule out the possibility that the trend is produced by a deficit of low-luminosity low-mass galaxies in protocluster environment. \\

\noindent - Our search for Ly$\alpha$ nebulae in the entire field yields a single  nebula with the total Ly$\alpha$ luminosity $\approx 2\times 10^{43}$~erg~s$^{-1}$ and the half-light radius (assuming an exponentially declining profile) of at least 5\arcsec\ (39~kpc at $z=3.13$).  Its location at the outer edge of the LAE overdensity may support a physical picture advocated by \citet{Badescu17}, that Ly$\alpha$ nebulae  trace group-sized halos falling in towards the protocluster center. The large variations seen in the observed number of Ly$\alpha$ nebulae around protoclusters hint at the short-lived nature of the phenomenon, perhaps brought on by galaxy-galaxy interactions. \\

\noindent - We have also identified a  brightest cluster galaxy candidate located $\approx$2\arcmin\  from the center of the LBG overdensity. The galaxy is one of the brightest LBGs in our sample, and has already assembled a stellar mass of $\approx 2\times 10^{11}M_\odot$. A full SED modeling suggests that a highly dust-obscured AGN dominates its mid-infrared flux at $\lambda_{\rm obs} \gtrsim 8\mu$m while still active star formation is responsible for a fairly reddened rest-frame UV and optical part of its SED. While the AGN-driven quenching of star formation in an already massive cluster galaxy fits the general expectation of how and when cluster galaxies formed, further validation (i.e., high spatial resolution imaging and a spectroscopic redshift) is needed to determine whether it is physically associated with the galaxy overdensity.

\acknowledgments
We thank the anonymous referee for a careful reading of the manuscript and a thoughtful report. 
The authors thank Yujin Yang and Yi-Kuan Chiang for useful comments and discussions. NM acknowledges the support provided by the funding for the ByoPiC project from the European Research Council (ERC) under the European Union's Horizon 2020 research and innovation programme grant agreement ERC-2015-AdG 695561.
This work is based on observations at Kitt Peak National Observatory, National Optical Astronomy Observatory (NOAO Prop. ID 2016B-0087; PI: K.-S. Lee), which is operated by the Association of Universities for Research in Astronomy (AURA) under cooperative agreement with the National Science Foundation. The authors are honored to be permitted to conduct astronomical research on Iolkam DuÕag (Kitt Peak), a mountain with particular significance to the Tohono OÕodham. 
 
\bibliographystyle{apj}
\bibliography{apj-jour,refs}

\LongTables

\begin{deluxetable*}{ccccccccccc}
\tabletypesize{\scriptsize}
\tablewidth{0pt}
\tablecaption{Catalog of LAE candidates \label{tbl2}}
\tablehead{
\colhead{ID} & \colhead{R.A.} & \colhead{Decl.} & \colhead{log($L_{{\rm Ly} \alpha}$)} & \colhead{log($L_{1700}$)} & \colhead{$W_{0,{\rm Ly}\alpha}$} & \colhead{LBG\tablenotemark{$\star$}} & \colhead{$o3-g$} & \colhead{$g-r$} & $o3$ & $\beta$\\
\colhead{} & \colhead{(J2000)} & \colhead{(J2000)} & \colhead{[erg~s$^{-1}$]} & \colhead{[erg~s$^{-1}$ Hz$^{-1}$]} & \colhead{[\AA]} & \colhead{} & \colhead{} & \colhead{} &  & 
}

\startdata
LAE11899\tablenotemark{$\dagger$} & 36.331703 & -4.62293 & 42.27$\pm$0.14      & 28.20$\pm$0.09         & 20.1$\pm$8.3       & 1   & -1.05$\pm$0.13       & 0.52$\pm$0.06 & 24.68$\pm$0.19       & -1.8$\pm$0.2     \\
LAE31361\tablenotemark{$\dagger$} & 36.147567 & -4.34230  & 43.01$\pm$0.04      & 28.49$\pm$0.04         & 49.2$\pm$6.8       & 1   & -1.65$\pm$0.06       & 0.29$\pm$0.04    & 23.11$\pm$0.07 & $-1.8\pm0.4$  \\
LAE33092\tablenotemark{$\dagger$} & 36.133734 & -4.31694 & 42.46$\pm$0.11       & 28.10$\pm$0.09         & 35.6$\pm$12.8       & 1   & -1.40$\pm$0.14       & 0.43$\pm$0.09      & 24.40$\pm$0.19 & $-2.5\pm0.7$ \\
LAE34191\tablenotemark{$\dagger$} & 36.112236 & -4.30260  & 42.82$\pm$0.06      & 28.64$\pm$0.04         & 22.9$\pm$3.8       & 1   & -1.12$\pm$0.05       & 0.44$\pm$0.03  & 23.37$\pm$0.08 & $-2.1\pm0.5$ \\
QSO30046 & 36.461127 & -4.36170 & 43.87$\pm$0.00 & 29.49$\pm$0.00 & 81.6$\pm$1.3 & 0 &  -1.96$\pm$0.01 & 1.30$\pm$0.01 & 21.06$\pm$0.01 & $-3.9\pm0.5$ \\
LAB17139 & 36.082992 & -4.54653 & 43.32$\pm$0.03      & 28.62$\pm$0.03         & 63.3$\pm$7.3       & 0   & -1.81$\pm$0.05       & 0.13$\pm$0.05  & 22.37$\pm$0.06 & $-2.6\pm0.8$     \\
LAE1223  & 36.548107 & -4.78182 & 42.62$\pm$0.10      & 28.21$\pm$0.08         & 33.3$\pm$10.1      & 1   & -1.38$\pm$0.11       & 0.17$\pm$0.07  &  23.99$\pm$0.16 & $-1.6\pm0.1$    \\
LAE3003  & 36.544044 & -4.75342 & 42.42$\pm$0.11       & 28.00$\pm$0.10         & 37.6$\pm$13.3      & 1   & -1.48$\pm$0.15       & 0.18$\pm$0.12 & 24.52$\pm$0.18 & $-1.1\pm0.2$     \\
LAE3007  & 36.347336 & -4.75314 & (42.03)      & (27.09)         & (123.3)      & 1   & -1.65$\pm$0.19       & $<$-0.41  & 25.69$\pm$0.75 & (-1.7)    \\
LAE3622  & 36.559838 & -4.74334 & 42.18$\pm$0.15      & 27.74$\pm$0.13         & 44.3$\pm$22.4      & 1   & -1.59$\pm$0.16       & 0.33$\pm$0.14  & 25.18$\pm$0.28 & -1.2$\pm$0.1    \\
LAE3833  & 36.399597 & -4.74040  & 42.70$\pm$0.04      & 28.33$\pm$0.04         & 42.0$\pm$5.9       & 1   & -1.54$\pm$0.06       & 0.52$\pm$0.05   & 23.85$\pm$0.07 & -1.9$\pm$0.2   \\
LAE4290  & 36.580112 & -4.73362 & (42.56)      & (27.09)         & (482.1)      & 0   & $<$-4.30       & --    & 24.49$\pm$0.11 & (-1.7)  \\
LAE4423  & 36.532995 & -4.73160  & (42.21)      & (27.05)         & (130.1) & 1   & -2.33$\pm$0.26       & -0.46$\pm$0.59  & 25.29$\pm$0.19 & (-1.7)    \\
LAE4564  & 36.196192 & -4.73027 & 42.53$\pm$0.05      & 27.60$\pm$0.10         & 95.3$\pm$21.4       & 0   & -2.15$\pm$0.12       & -0.35$\pm$0.24   & 24.42$\pm$0.09 & -0.3$\pm$0.2   \\
LAE4614  & 36.531916 & -4.72931 & 42.38$\pm$0.10      & 28.15$\pm$0.09         & 29.6$\pm$9.7        & 1  & -1.31$\pm$0.16       & 0.47$\pm$0.10 & 24.56$\pm$0.15       & -1.3$\pm$0.5       \\
LAE5378  & 36.626309 & -4.71876 & 42.41$\pm$0.10       & 27.45$\pm$0.13         & 174.8$\pm$93.0      & 1   & -2.51$\pm$0.21       & 0.24$\pm$0.24    & 24.78$\pm$0.22       & -2.0$\pm$1.7  \\
LAE7142  & 36.383586 & -4.69378 & 42.41$\pm$0.12      & 28.05$\pm$0.10         & 35.5$\pm$13.6      & 1   & -1.43$\pm$0.16       & 0.32$\pm$0.10   & 24.52$\pm$0.19       & -1.4$\pm$0.2    \\
LAE7560  & 36.517002 & -4.68750  & 42.32$\pm$0.07      & 27.68$\pm$0.08         & 48.4$\pm$12.4       & 1   & -1.61$\pm$0.12       & 0.06$\pm$0.11 & 24.83$\pm$0.13       & -2.6$\pm$0.5     \\
LAE8382  & 36.475010  & -4.67657 & 42.77$\pm$0.04      & 28.35$\pm$0.04         & 34.4$\pm$5.0       & 1   & -1.39$\pm$0.05       & 0.21$\pm$0.04  & 23.64$\pm$0.07       & -1.9$\pm$0.1    \\
LAE8783  & 36.262082 & -4.67020  & 42.37$\pm$0.06      & 27.41$\pm$0.10         & 106.5$\pm$29.2      & 1   & -2.21$\pm$0.14       & -0.22$\pm$0.24   & 24.83$\pm$0.12       & -1.1$\pm$0.4   \\
LAE9355  & 36.224172 & -4.66214 & 42.78$\pm$0.03      & 27.99$\pm$0.03         & 87.3$\pm$11.6      & 1   & -2.07$\pm$0.07       & 0.08$\pm$0.09 & 23.79$\pm$0.06       & -1.4$\pm$0.7     \\
LAE9436  & 36.416560  & -4.66001 & 42.11$\pm$0.13      & 27.28$\pm$0.17         & 52.5$\pm$25.1      & 1   & -1.72$\pm$0.19       & -0.52$\pm$0.32 & 25.38$\pm$0.25       & -0.8$\pm$2.3     \\
LAE9589  & 36.276072 & -4.65831 & 42.60$\pm$0.06      & 27.95$\pm$0.06         & 48.2$\pm$10.4       & 1   & -1.63$\pm$0.08       & -0.05$\pm$0.10 & 24.14$\pm$0.11       & -1.9$\pm$1.0      \\
LAE9596  & 36.207723 & -4.65882 & 42.79$\pm$0.04      & 28.27$\pm$0.02         & 47.1$\pm$6.4       & 1   & -1.61$\pm$0.06       & 0.24$\pm$0.05 & 23.65$\pm$0.07       & -2.0$\pm$0.2     \\
LAE11088 & 36.513041 & -4.63546 & 42.34$\pm$0.11       & 27.62$\pm$0.12         & 71.0$\pm$30.2      & 1   & -1.89$\pm$0.18       & 0.28$\pm$0.17 & 24.97$\pm$0.22       & -2.4$\pm$0.8     \\
LAE11215 & 36.449218 & -4.63364 & 42.26$\pm$0.08      & 27.26$\pm$0.14         & 167.9$\pm$81.2      & 1   & -2.50$\pm$0.23       & 0.09$\pm$0.34 & 25.15$\pm$0.18       & -1.7$\pm$2.4     \\
LAE11424 & 36.531781 & -4.63014 & 42.29$\pm$0.16      & 28.18$\pm$0.10         & 18.8$\pm$8.3        & 1   & -1.00$\pm$0.12       & 0.39$\pm$0.05 & 24.61$\pm$0.21       & -2.0$\pm$0.4     \\
LAE11535 & 36.066720  & -4.62925 & 42.89$\pm$0.07      & 28.87$\pm$0.06         & 33.1$\pm$7.8       & 0   & -1.42$\pm$0.09       & 0.89$\pm$0.04  & 23.34$\pm$0.12 & 0.5$\pm$0.5     \\
LAE11926 & 36.472781 & -4.62227 & 42.31$\pm$0.07      & 27.60$\pm$0.10         & 95.7$\pm$29.5       & 1   & -2.17$\pm$0.15       & 0.03$\pm$0.19 & 24.97$\pm$0.14       & 1.5$\pm$0.3     \\
LAE12763 & 36.481400   & -4.60974 & 42.72$\pm$0.04      & 27.87$\pm$0.05         & 127.9$\pm$26.0      & 0   & -2.31$\pm$0.08       & 0.31$\pm$0.09 & 23.99$\pm$0.09       & -2.2$\pm$0.3     \\
LAE12872 & 36.438421 & -4.60755 & 42.46$\pm$0.09      & 27.99$\pm$0.08         & 30.6$\pm$9.9       & 1   & -1.29$\pm$0.12       & 0.19$\pm$0.08 & 24.53$\pm$0.17       & -2.5$\pm$0.1     \\
LAE12900 & 36.049683 & -4.60644 & 42.22$\pm$0.10      & 27.04$\pm$0.23         & 227.1$\pm$206.1    & 1   & -2.80$\pm$0.29       & -0.26$\pm$0.58    & 25.27$\pm$0.21 & -0.1$\pm$3.6   \\
LAE13337 & 36.189058 & -4.60115 & (42.33)      & (27.21)          & (150.2)      & 1   & -2.43$\pm$0.21       & -0.29$\pm$0.41    & 24.97$\pm$0.13 & (-1.7)   \\
LAE13519 & 36.344731 & -4.59843 & 42.57$\pm$0.08      & 28.05$\pm$0.07         & 46.9$\pm$12.8       & 1   & -1.62$\pm$0.11       & 0.23$\pm$0.09   & 24.21$\pm$0.14 & -1.7$\pm$0.7    \\
LAE13642 & 36.256098 & -4.59568 & (42.16)      & (27.12)         & (144.9)       & 1   & -2.40$\pm$0.25       & -0.13$\pm$0.42    & 25.38$\pm$0.16 & (-1.7)   \\
LAE13732 & 36.281673 & -4.59482 & 42.29$\pm$0.10       & 28.53$\pm$0.07         & 23.3$\pm$7.0       & 0   & -1.16$\pm$0.13       & 1.39$\pm$0.04 & 24.70$\pm$0.14 & -1.2$\pm$0.5      \\
LAE14869 & 36.431008 & -4.57703 & 42.61$\pm$0.08      & 28.15$\pm$0.08         & 48.5$\pm$14.5      & 1   & -1.72$\pm$0.12       & 0.37$\pm$0.09   & 24.05$\pm$0.16 & -1.3$\pm$0.1    \\
LAE16775 & 36.263772 & -4.54861 & 42.19$\pm$0.15      & 27.96$\pm$0.11         & 22.7$\pm$10.7       & 1   & -1.12$\pm$0.16       & 0.27$\pm$0.09 & 24.94$\pm$0.23 & -1.8$\pm$0.1      \\
LAE17045 & 36.598313 & -4.54491 & 42.43$\pm$0.06      & 27.63$\pm$0.09         & 79.5$\pm$20.9      & 1   & -1.99$\pm$0.14       & 0.00$\pm$0.16 & 24.65$\pm$0.12 & -1.7$\pm$1.2      \\
LAE17728 & 36.171767 & -4.53675 & 42.80$\pm$0.06      & 28.26$\pm$0.05         & 45.2$\pm$9.3       & 1   & -1.57$\pm$0.07       & 0.22$\pm$0.05 & 23.62$\pm$0.11 & -2.4$\pm$0.1      \\
LAE18151 & 36.253813 & -4.52938 & 42.48$\pm$0.11       & 28.24$\pm$0.08         & 21.5$\pm$7.2       & 1   & -1.07$\pm$0.10        & 0.24$\pm$0.05 & 24.20$\pm$0.16 & -2.4$\pm$0.1      \\
LAE18455 & 36.221296 & -4.52418 & 42.14$\pm$0.08      & 27.23$\pm$0.12         & 150.2$\pm$68.3      & 0   & -2.40$\pm$0.22       & 0.35$\pm$0.28 & 25.45$\pm$0.18 & -2.9$\pm$2.2      \\
LAE18514 & 36.153655 & -4.52340  & 42.27$\pm$0.10      & 27.76$\pm$0.11         & 63.5$\pm$24.7      & 1   & -1.87$\pm$0.18       & 0.31$\pm$0.17 & 25.00$\pm$0.19 & 0.2$\pm$0.1      \\
LAE19027 & 36.069764 & -4.51725 & 42.65$\pm$0.05      & 28.06$\pm$0.05         & 55.7$\pm$10.0       & 1   & -1.74$\pm$0.08       & 0.23$\pm$0.07    & 24.04$\pm$0.09 & -1.7$\pm$0.1   \\
LAE19205 & 36.177991 & -4.51562 & 42.85$\pm$0.06      & 28.64$\pm$0.05         & 33.6$\pm$6.0       & 1   & -1.39$\pm$0.07       & 0.64$\pm$0.04 & 23.41$\pm$0.10 & -1.4$\pm$0.4      \\
LAE19465 & 36.190375 & -4.51029 & 42.53$\pm$0.09      & 28.47$\pm$0.08         & 70.6$\pm$22.0      & 0   & -1.92$\pm$0.11       & 1.72$\pm$0.05   & 24.38$\pm$0.17 & -1.4$\pm$0.4    \\
LAE19469 & 36.103684 & -4.51081 & 42.86$\pm$0.05      & 28.56$\pm$0.04         & 34.1$\pm$5.8       & 1   & -1.40$\pm$0.06       & 0.47$\pm$0.04  & 23.39$\pm$0.09 & -1.7$\pm$0.4     \\
LAE19503 & 36.103966 & -4.50951 & 42.53$\pm$0.06      & 28.22$\pm$0.06         & 31.7$\pm$6.9       & 1   & -1.34$\pm$0.11       & 0.42$\pm$0.06  & 24.21$\pm$0.10 & -2.0$\pm$0.4     \\
LAE19965 & 36.570704 & -4.50256 & 42.31$\pm$0.16      & 28.13$\pm$0.11         & 26.0$\pm$12.5       & 1   & -1.22$\pm$0.13       & 0.48$\pm$0.06   & 24.69$\pm$0.25 & -1.5$\pm$0.1    \\
LAE20057 & 36.184999 & -4.50057 & (42.32)      & (27.09)         & (427.9)    & 0   & -2.98$\pm$0.33       & $<$-0.37   & 25.05$\pm$0.18 & (-1.7)    \\
LAE20136 & 36.169522 & -4.49970  & 42.32$\pm$0.11      & 27.67$\pm$0.12         & 59.6$\pm$24.5      & 1   & -1.79$\pm$0.16       & 0.11$\pm$0.16    & 24.87$\pm$0.22 & -1.6$\pm$1.1   \\
LAE20210 & 36.184859 & -4.49883 & 42.34$\pm$0.07      & 27.59$\pm$0.10         & 96.0$\pm$31.9      & 1   & -2.13$\pm$0.16       & 0.26$\pm$0.17   & 24.90$\pm$0.15 & -1.6$\pm$1.0    \\
LAE20309 & 36.272384 & -4.49789 & 42.52$\pm$0.10      & 28.69$\pm$0.07         & 22.8$\pm$6.8       & 1   & -1.15$\pm$0.12       & 1.17$\pm$0.04  & 24.11$\pm$0.14 & -1.1$\pm$0.3     \\
LAE20334 & 36.210326 & -4.49665 & 42.30$\pm$0.12      & 27.76$\pm$0.11         & 49.3$\pm$21.0       & 1   & -1.63$\pm$0.15       & 0.30$\pm$0.12  & 24.89$\pm$0.23 & -2.3$\pm$0.3     \\
LAE20391 & 36.574596 & -4.49574 & 42.22$\pm$0.11       & 27.89$\pm$0.09         & 31.1$\pm$11.2       & 1   & -1.32$\pm$0.15       & 0.35$\pm$0.09  & 24.97$\pm$0.18 & -2.2$\pm$0.8     \\
LAE21289 & 36.075996 & -4.48283 & 42.40$\pm$0.06      & 27.51$\pm$0.10         & 114.1$\pm$37.1      & 1   & -2.25$\pm$0.16       & 0.08$\pm$0.19  & 24.78$\pm$0.14 & -1.7$\pm$0.1     \\
LAE21315 & 36.078332 & -4.48304 & 42.62$\pm$0.07      & 27.96$\pm$0.09         & 87.5$\pm$27.6      & 1   & -2.09$\pm$0.14       & 0.34$\pm$0.14  & 24.19$\pm$0.15 & -0.7$\pm$0.1     \\
LAE21887 & 36.143396 & -4.47564 & 42.94$\pm$0.05      & 28.29$\pm$0.05         & 102.8$\pm$20.7      & 1   & -2.18$\pm$0.08       & 0.58$\pm$0.07    & 23.41$\pm$0.10 & -1.5$\pm$1.2   \\
LAE21996 & 36.137965 & -4.47260  & 42.67$\pm$0.08      & 28.07$\pm$0.08         & 59.4$\pm$17.7       & 1   & -1.75$\pm$0.11       & 0.36$\pm$0.09  & 24.00$\pm$0.16 & -2.8$\pm$1.2     \\
LAE22629 & 36.470213 & -4.46346 & 42.34$\pm$0.09      & 27.93$\pm$0.08         & 49.0$\pm$15.7       & 1   & -1.66$\pm$0.12       & 0.48$\pm$0.09  & 24.78$\pm$0.17 & -1.3$\pm$0.8     \\
LAE23293 & 36.058462 & -4.45461 & 43.13$\pm$0.02      & 28.39$\pm$0.03         & 73.4$\pm$7.2       & 1   & -1.94$\pm$0.04       & 0.09$\pm$0.04    & 22.88$\pm$0.05 & -1.8$\pm$0.3   \\
LAE23302 & 36.337272 & -4.45334 & 42.25$\pm$0.08      & 27.21$\pm$0.16         & 132.0$\pm$60.0      & 0   & -2.33$\pm$0.22       & -0.14$\pm$0.35  & 25.15$\pm$0.18 & -2.1$\pm$2.8     \\
LAE23320 & 36.057467 & -4.45327 & 42.37$\pm$0.09      & 27.52$\pm$0.13         & 117.0$\pm$52.2      & 1   & -2.28$\pm$0.19       & 0.12$\pm$0.23    & 24.85$\pm$0.20 & -0.9$\pm$0.1   \\
LAE24442 & 36.133481 & -4.43661 & 42.25$\pm$0.12       & 27.97$\pm$0.08         & 20.9$\pm$7.1       & 1   & -1.04$\pm$0.13       & 0.10$\pm$0.07  & 24.75$\pm$0.15 & -2.5$\pm$1.0     \\
LAE25508 & 36.105297 & -4.42287 & 42.52$\pm$0.11       & 28.05$\pm$0.10         & 59.1$\pm$23.3      & 1   & -1.80$\pm$0.16       & 0.50$\pm$0.13   & 24.37$\pm$0.21 & -1.1$\pm$0.4    \\
LAE26131 & 36.476590  & -4.41338 & 42.12$\pm$0.12      & 27.73$\pm$0.11         & 32.4$\pm$13.3      & 1   & -1.35$\pm$0.15       & 0.23$\pm$0.10   & 25.23$\pm$0.21 & -2.0$\pm$1.7     \\
LAE26308 & 36.521510  & -4.41145 & 42.31$\pm$0.13      & 27.24$\pm$0.19         & 223.7$\pm$170.5     & 1   & -2.68$\pm$0.24       & -0.04$\pm$0.41   & 25.06$\pm$0.30 & -0.4$\pm$0.1    \\
LAE26947 & 36.401936 & -4.40294 & 42.36$\pm$0.08      & 27.69$\pm$0.08         & 38.9$\pm$10.7       & 1   & -1.45$\pm$0.09       & -0.20$\pm$0.10   & 24.69$\pm$0.15 & -2.9$\pm$0.1     \\
LAE28041 & 36.147012 & -4.38820  & 42.61$\pm$0.10      & 28.29$\pm$0.08         & 31.2$\pm$9.6       & 1   & -1.33$\pm$0.11       & 0.36$\pm$0.06   & 24.00$\pm$0.16 & -1.9$\pm$0.5    \\
LAE28534 & 36.261246 & -4.38168 & 43.06$\pm$0.03      & 28.53$\pm$0.03        & 52.1$\pm$5.9       & 1   & -1.70$\pm$0.05       & 0.25$\pm$0.04   & 22.99$\pm$0.06 & -1.3$\pm$0.2    \\
LAE28985 & 36.275809 & -4.37465 & 42.48$\pm$0.06      & 27.72$\pm$0.08         & 76.2$\pm$20.2       & 1   & -1.94$\pm$0.12       & 0.13$\pm$0.13  & 24.51$\pm$0.13 & -2.5$\pm$1.0     \\
LAE29112 & 36.574609 & -4.37290  & (42.50)      & (27.09)         & (417.3)    & 0   & $<$-4.12       & -- & 24.68$\pm$0.18 & (-1.7)       \\
LAE30621 & 36.335944 & -4.35250  & 42.41$\pm$0.09      & 28.25$\pm$0.07         & 20.5$\pm$6.0       & 1   & -1.09$\pm$0.10        & 0.23$\pm$0.06  & 24.36$\pm$0.14 & -0.8$\pm$0.6     \\
LAE31356 & 36.378076 & -4.34067 & 42.46$\pm$0.08      & 27.72$\pm$0.11         & 68.3$\pm$22.5      & 1   & -1.90$\pm$0.17       & -0.01$\pm$0.20   & 24.55$\pm$0.15 & -1.4$\pm$0.9     \\
LAE31802 & 36.128128 & -4.33464 & (42.20)      & (27.09)         & (212.3) & 1   & $<$-3.45       & -- & 25.34$\pm$0.17 & (-1.7)       \\
LAE32293 & 36.376859 & -4.32791 & 42.13$\pm$0.15      & 28.02$\pm$0.11         & 24.7$\pm$11.1       & 0   & -1.19$\pm$0.16       & 0.57$\pm$0.08  & 25.11$\pm$0.22 & -1.3$\pm$0.1     \\
LAE33330 & 36.425224 & -4.31357 & 42.41$\pm$0.10       & 27.71$\pm$0.12         & 75.3$\pm$30.5      & 1   & -1.89$\pm$0.20        & 0.40$\pm$0.19  & 24.69$\pm$0.20 & -3.6$\pm$1.6     \\
LAE34533 & 36.392224 & -4.29760  & 42.56$\pm$0.07      & 28.02$\pm$0.07         & 49.0$\pm$12.5      & 1   & -1.66$\pm$0.10        & 0.19$\pm$0.09  & 24.24$\pm$0.14 & -1.4$\pm$0.6     \\
LAE35011 & 36.071438 & -4.28969 & 42.33$\pm$0.13      & 28.07$\pm$0.10         & 23.6$\pm$9.5       & 1   & -1.15$\pm$0.14       & 0.22$\pm$0.07  & 24.60$\pm$0.19 & -1.8$\pm$0.1     \\
LAE35344 & 36.578088 & -4.28621 & 42.19$\pm$0.44      & 27.80$\pm$0.33         & 49.7$\pm$68.5      & 1   & -1.03$\pm$0.14       & 0.61$\pm$0.06   & 25.17$\pm$0.82 & -1.6$\pm$0.9    \\
LAE35637 & 36.360586 & -4.28129 & 42.07$\pm$0.14      & 27.99$\pm$0.09         & 18.3$\pm$7.5        & 1   & -0.99$\pm$0.15       & 0.44$\pm$0.07  & 25.15$\pm$0.18 & -1.8$\pm$0.5     \\
LAE35739 & 36.566167 & -4.27998 & 42.39$\pm$0.11       & 27.93$\pm$0.11          & 66.0$\pm$26.6     & 0   & -1.87$\pm$0.17       & 0.68$\pm$0.12  & 24.73$\pm$0.21 & -1.4$\pm$0.4     \\
LAE35993 & 36.263103 & -4.27687 & 42.59$\pm$0.06      & 27.32$\pm$0.16         & 194.9$\pm$71.6     & 1   & -2.59$\pm$0.16       & -0.53$\pm$0.40    & 24.34$\pm$0.13 & -1.4$\pm$0.1    \\
LAE36621 & 36.504879 & -4.26776 & 42.12$\pm$0.16      & 28.11$\pm$0.10         & 17.5$\pm$8.0       & 1   & -0.97$\pm$0.15       & 0.52$\pm$0.07  & 25.00$\pm$0.20 & -1.7$\pm$0.3     \\
LAE36658 & 36.557478 & -4.26803 & 42.55$\pm$0.08      & 27.54$\pm$0.14         & 311.4$\pm$206.0     & 1   & -2.86$\pm$0.23       & 0.19$\pm$0.36  & 24.47$\pm$0.19 & 0.3$\pm$1.9     \\
LAE37505 & 36.255731 & -4.25733 & 42.22$\pm$0.13      & 27.81$\pm$0.10         & 33.9$\pm$14.1      & 1   & -1.40$\pm$0.11       & 0.17$\pm$0.09  & 25.00$\pm$0.23 & -1.5$\pm$0.3    \\
LAE37991 & 36.500602 & -4.25212 & 42.76$\pm$0.04      & 27.89$\pm$0.06         & 87.3$\pm$14.6      & 1   & -2.05$\pm$0.09       & -0.03$\pm$0.14  & 23.84$\pm$0.07 & -2.3$\pm$0.4     \\
LAE38096 & 36.625888 & -4.24970  & 42.41$\pm$0.07      & 27.48$\pm$0.12         & 125.8$\pm$47.7      & 1   & -2.33$\pm$0.17       & -0.04$\pm$0.26   & 24.76$\pm$0.16 & -0.8$\pm$0.3    \\
LAE38409 & 36.533597 & -4.24540  & 42.34$\pm$0.06      & 27.29$\pm$0.15         & 320.2$\pm$206.4     & 1   & -2.86$\pm$0.26       & 0.27$\pm$0.41  & 25.00$\pm$0.14 & -0.9$\pm$0.8     \\
LAE38991 & 36.598963 & -4.23799 & 42.34$\pm$0.07      & 27.38$\pm$0.13         & 108.0$\pm$36.8      & 1   & -2.19$\pm$0.18       & -0.09$\pm$0.30  & 24.91$\pm$0.14 & -2.2$\pm$2.4      \\
LAE40397 & 36.449092 & -4.21805 & 42.17$\pm$0.14      & 27.79$\pm$0.11         & 24.7$\pm$11.0      & 1   & -1.15$\pm$0.16       & 0.02$\pm$0.12    & 25.02$\pm$0.22 & -2.4$\pm$0.9   \\
LAE40773 & 36.072466 & -4.21316 & 42.73$\pm$0.06      & 27.90$\pm$0.07         & 79.1$\pm$18.4      & 1   & -2.01$\pm$0.09       & -0.16$\pm$0.12 & 23.89$\pm$0.12 & -0.9$\pm$1.4      \\
LAE41503 & 36.088564 & -4.20889 & 42.93$\pm$0.04      & 28.35$\pm$0.04         & 53.5$\pm$7.1       & 1   & -1.72$\pm$0.06       & 0.20$\pm$0.04   & 23.34$\pm$0.07 & -1.5$\pm$0.4    \\
LAE42838 & 36.164882 & -4.20061 & (42.80)      & (27.31)         & (944.1)   & 1   & -3.23$\pm$0.32       & -0.33$\pm$0.72   & 23.88$\pm$0.16 & (-1.7)    
\enddata

\tablenotetext{$\dagger$}{The spectroscopic redshifts are $z$=3.133, 3.124, 3.131, and 3.130 for LAE11899, LAE31361, LAE33092, and LAE34191, respectively.}
\tablenotetext{$\star$}{Whether this LAE also satisfies the LBG selection criteria; 1--yes, 0--no.}

\tablecomments{In the cases where no significant continuum flux is detected, we use 2$\sigma$ flux limit for calculating the colors. For sources where there are no $\beta$ measurements (i.e., at least two broad bands are not detected), we use the median $\beta$ value of the other sources, and give an estimate in parentheses. For sources not detected in $g$, a 3$\sigma$ flux limit is used to estimate the continuum flux and equivalent width.}

\end{deluxetable*}

\end{document}